\documentclass[twocolumn]{aastex631}
\usepackage[normalem]{ulem}
\usepackage{makecell}
\usepackage{graphics,xcolor}
\usepackage{amsmath}
\usepackage{soul}
\usepackage{CJK}
\usepackage{listings}
\usepackage{comment}

\newcommand{\halpha}{{\rm H}\alpha}
\newcommand{\haew}{{\rm H}\alpha \; {\rm EW}}
\newcommand{\lhalbol}{L_{{\rm H}\alpha}/L_{\rm bol}}
\newcommand{\pseudoha}{{\rm H}\alpha{\rm \,pEW}}

\newcommand{\Msun}{\mbox{${\rm M}_{\sun}$}}
\newcommand{\Rsun}{\mbox{${\rm R}_{\sun}$}}

\begin{document}

\title{Accurate and Model Independent Radius Determination of Single FGK and M Dwarfs Using Gaia~DR3 Data}
\shorttitle{radius of m dwarfs}
\shortauthors{Kiman et al.}
 
\received{22-Jan-2024}
\revised{22-Apr-2024}
\accepted{20-May-2024}
\submitjournal{Astronomical Journal}

\author[0000-0003-2102-3159]{Rocio Kiman}
\affil{Department of Astronomy, California Institute of Technology, Pasadena, CA 91125, USA}
\affil{Kavli Institute for Theoretical Physics, University of California, Santa Barbara, CA 93106, USA}
\correspondingauthor{Rocio Kiman}
\email{rociokiman@gmail.com}

\author[0000-0003-2630-8073]{Timothy D. Brandt}
\affil{Space Telescope Science Institute, 3700 San Martin Dr, Baltimore, MD 21218, USA}
\affil{Department of Physics, University of California, Santa Barbara, Santa Barbara, CA 93106, USA}

\author[0000-0001-6251-0573]{Jacqueline K. Faherty}
\affil{Department of Astrophysics, American Museum of Natural History, Central Park West at 79th St, New York, NY 10024, USA}

\author[0000-0001-9482-7794]{Mark Popinchalk}
\affil{Department of Astrophysics, American Museum of Natural History, Central Park West at 79th St, New York, NY 10024, USA}
\affil{Physics, The Graduate Center, City University of New York, New York, NY 10016, USA}
\affil{Department of Physics and Astronomy, Hunter College, City University of New York, 695 Park Avenue, New York, NY 10065, USA}

\begin{abstract} 

Measuring fundamental stellar parameters is key to fully comprehending the evolution of stars. However, current theoretical models over-predict effective temperatures, and under-predict radii, compared to observations of K and M dwarfs (radius inflation problem). In this work, we developed a model independent method to infer precise radii of single FGK and M dwarfs using Gaia~DR3 parallaxes and photometry, and we used it to study the radius inflation problem. We calibrated nine surface brightness-color relations for the three Gaia magnitudes and colors using a sample of stars with angular diameter measurements. We achieved an accuracy of $4\%$ in our angular diameter estimations, which Gaia's parallaxes allow us to convert to a physical radii. We validated our method by comparing our radius measurements with literature samples and the Gaia~DR3 catalog, which confirmed the accuracy of our method and revealed systematic offsets in the Gaia measurements. Moreover, we used a sample with measured $\halpha$ equivalent width ($\haew$), a magnetic activity indicator, to study the radius inflation problem. We demonstrated that active stars have larger radii than inactive stars, showing that radius inflation is correlated with magnetic activity. We found a correlation between the radius inflation of active stars and $\haew$ for the mass bin $0.5<M[{\rm M_\odot}]\leq 0.6$, but we found no correlation for lower masses. This could be due to lack of precision in our radius estimation or a physical reason. Radius measurements with smaller uncertainties are necessary to distinguish between the two scenarios. 

\end{abstract}

\section{Introduction}
\label{sec:intro}

Measuring fundamental properties of stars, like radius and effective temperature, is key to understanding stellar evolution and developing precise theoretical models.
Currently there is disagreement between models and measurements of radius and effective temperatures of low-mass stars. 
Measured radii are larger than what models predict for their mass, age and metallicity, and effective temperatures are lower. 
This phenomenon is known as radius inflation, and since the first time it was discovered \citep{Hoxie1970,Hoxie1973}, the difference between models and measurements for radius and temperature has been studied extensively.
Inflation has been calibrated by several studies using both single and binary stars with results such as: $5-15\%$ inflation of the radius \citep{Ribas2006}, $15-20\%$ \citep{Berger2006} and $5-10\%$ \citep{Torres2013}, but without precise agreement.
The cause of radius inflation is still being debated, though the leading explanation is magnetic activity.

Magnetic activity has been shown theoretically to inhibit convective transport of energy, which generates an increase in the stellar radius, and a decrease of surface temperature. This effect could be caused by: strong magnetic fields \citep{Gough1966,Mullan2001,Chabrier2007,MacDonald2012,Feiden2013} or spot coverage \citep{Chabrier2007,Somers2015b,Somers2020}. 
Although in theory magnetic activity can explain the difference between radius measurements and models, not all observational studies agree. 
Several studies have found an empirical correlation between radius inflation and magnetic activity by using eclipsing binaries \citep{Lopez-Morales2007,Kraus2011,Irwin2011b,Feiden2012a,Birkby2012,Torres2018,Torres2019,Torres2020,Torres2021a}. Detached eclipsing binaries provide an optimal stage to study radius inflation because they allow for a precise measurement of the mass and radius of both components (${\sim}3\%$ uncertainty, \citealt{Torres2010}). 
Nonetheless, these studies do not confirm the relation between magnetic activity and inflation for single stars, given that the influence of tidal interactions make the stars in eclipsing binaries spin up, as it was pointed out in \citet{Torres2021a}. In this study the authors say that the rotation period of the binary is more than an order of magnitude faster than single stars of the same mass in the cluster \citep{Curtis2020}.

Previous studies have shown empirically that there is a correlation between radius inflation and magnetic activity for single stars \citep{Morales2008,Stassun2012}.
On the other hand, several studies have failed to find a correlation between radius inflation and magnetic activity or rotation for single stars \citep{Lopez-Morales2007,Mann2015,Kesseli2018,Morrell2019}, which could be due to activity being weaker in single stars in comparison to binary stars \citep{Demory2009,Boyajian2012b,Spada2013}.
Although the relation between magnetic activity and radius inflation for single stars has been studied extensively, a clear empirical correlation using direct estimation of the radius of the stars is still missing.

As mentioned above, eclipsing binaries have been used extensively in the literature to measure radius and mass with high precision, although most of these studies analyse a small number of stars due to the complexity of these measurements \citep[e.g., ][]{Torres2002,Lopez-Morales2005,Torres2014,Torres2018,Cruz2018,Torres2020,Torres2021a,Torres2021b,Jennings2023}. 
There have been recent studies with large samples of eclipsing binaries, but they have not achieved the same precision in radius and mass measurements as previous studies \citep{Mowlavi2022,Cruz2022}.
Another way to perform a precise analysis of radius inflation would be using a large enough sample to average out statistical uncertainty. In order to apply such a technique, we need a method to estimate radii which can be applied to a large sample of stars.

There are different ways to estimate a stellar radius for single stars, but obtaining an angular diameter measurement is the method that has comparable, or slightly larger, uncertainty to eclipsing binaries \citep{Spada2013}.
The Center for High Angular Resolution Astronomy (CHARA) Array \citep{tenBrummelaar2005,McAlister2005} is key for interferometric measurements of stellar angular diameters.
CHARA is an optical/IR interferometer, with six $1$\,m aperture telescopes in the shape of a Y, two telescopes in each branch, providing 15 baselines ranging from $34.1$ to $330.7$\,m. 
CHARA allows precise measurements of angular diameter for stars \citep[$\lesssim 5\%$, e.g., ][]{Berger2006,Boyajian2012a,Boyajian2012b}.
These measurements are observationally challenging and they require bright stars ($H, K' < 6.5$, $V < 11$, \citealt{Boyajian2012b}). 
Luckily, using a set of measurements of angular diameters, it is possible to estimate each star's surface brightness and calibrate the surface brightness-colour relation \citep[SBCR, ][]{Wesselink1969,Barnes1976,Barnes1976b,Fouque1997}, which allows the estimation of the angular diameter from the color and magnitude of a star.
This relation has been calibrated extensively in the literature using angular diameter measurements \citep[e.g., ][]{Kervella2004, Boyajian2014, Challouf2014, Salsi2020, Salsi2021} and using eclipsing binaries \citep[e.g., ][]{Graczyk2017}. 

With the release of the Gaia~DR3 all-sky survey of stellar astrometry and photometry \citep{Gaia2016,Gaia2022,Babusiaux2022}, it is now possible to calibrate the SBCR with unprecedented precision.
Moreover, combining the parallax measurements from Gaia with the SBCR, we can estimate stellar radius for most of the stars in the Gaia catalog. 
Recently \citet{Graczyk2021} used Gaia photometry to obtain the SBCR using eclipsing binaries, however their relation does not account for the M~dwarf regime, which is the main focus of this article.
\citet{Salsi2020,Salsi2021} calibrated the SBCR using a sample of angular diameter measurements from the literature \citep{Duvert2016} for F, G, K and M dwarfs. 
However, they fit the relation for FGK and M main sequence stars separately, which causes a systematic error when estimating the radius of stars, as we will show in this work.
In addition, they calibrated the relation for the $(G-K_{\rm s})$ color, causing the calculation of the angular diameter to depend on having magnitudes from The Two Micron All Sky Survey \citep[2MASS,][]{Skrutskie2006}.
Therefore, we decided to calibrate the SBCR for FGK and M dwarfs using only Gaia photometry, which will provide a more precise radius estimation.

In this study we calibrate the SBCR using only Gaia photometry. We use a sample of literature measurements of angular diameters, and we take advantage of the quality cuts of the Gaia data to select the best possible calibration sample and fit the relation.
The description of the calibration sample and the fit of the Gaia SBCR is in Section~\ref{sec:sbcr}. 
In Section~\ref{sec:comparison_literature} we use our Gaia SBCR to estimate radius for several literature samples with measured radius to analyze the performance of our method.
In Section~\ref{sec:biases} we study the influence of metallicity, extinction and variability on the estimation of radius with our method. 
Finally, in Section~\ref{sec:radiusinflation} we estimate the radii of a statistically large sample of M~dwarfs from \citet{Kiman2019} with $\halpha$ equivalent width measurements to study the relation between magnetic activity and radius inflation for single stars.




\section{Gaia~DR3 Surface Brightness-Colour Relation}
\label{sec:sbcr}

We started with the Stefan-Boltzmann law to derive the Gaia SBCR which will allow the measurement of stellar radii from photometry and parallaxes. 
To estimate the bolometric flux that we measure on Earth, we need to take into account the angular diameter ($\theta$) of the star, which transforms the Stefan-Boltzmann law to:

\begin{equation}
    F_{\rm BOL_{Earth}} = \frac{\theta^2}{4}\sigma T_{\rm eff}^4. \label{eq:boltzearth}
\end{equation}

\noindent where $\sigma$ is the Stefan-Boltzmann constant and $T_{\rm eff}$ is the effective temperature of the star. Equation~\eqref{eq:boltzearth} is from \citet{Casagrande2006}. The bolometric flux, $F_{\rm BOL_{Earth}}$, can be estimated as the flux in a photometric band $m_\lambda$ plus a bolometric correction which can be estimated as a polynomial of a color index~($X$) 

\begin{equation}
    F_{\rm BOL_{Earth}} \approx 10^{-0.4 m_\lambda} p_1(X). \label{eq:approx1}
\end{equation}

\noindent Using a similar approximation for the effective temperature, $T_{\rm eff}^4$, we get

\begin{equation}
    T_{\rm eff}^4 \approx p_2 (X) \label{eq:approx2}
\end{equation}

\noindent where $p_2(X)$ is a polynomial of the color. Combining equations \eqref{eq:boltzearth} and the approximations (Equations \eqref{eq:approx1} and \eqref{eq:approx2}) we obtain  

\begin{equation}
    \theta ^2 \approx 10^{-0.4 m_\lambda}p_3(X)  \label{eq:SBCR}
\end{equation}

\noindent where $p_3(X) = p_1(X)/p_2(X)$. Finally, for our calibration we define $S_{\rm m_\lambda}$ such that 

\begin{equation}
    \log _{10} S_{\rm m_\lambda} = \log_{10}(\theta ^2 10^{0.4 m_\lambda}) = p(X) \label{eq:SBCR_flux}
\end{equation}

\noindent for a magnitude $m_\lambda$ and a color index $X$.
Equation~\eqref{eq:SBCR_flux} is our SBCR. 
In the rest of this section we will calibrate Equation~\eqref{eq:SBCR_flux} for the three Gaia bands $m_\lambda = G$, $G_{\rm RP}$, $G_{\rm BP}$, and the three Gaia colors $X = (G_{\rm BP}-G_{\rm RP})$, $(G-G_{\rm RP})$, $(G_{\rm BP}-G)$.





\subsection{Calibration sample}
\label{subsec:calibrationsample}

We calibrated the SBCR described in Equation~\eqref{eq:SBCR_flux} using the catalog JMMC (Jean-Marie Mariotti Center) Measured Stellar Diameters Catalog \citep[JMDC, ][]{Duvert2016}, which is a compilation of angular diameter measurements for stars of different masses and evolutionary stages. 
The catalog has $2013$ rows, with $1061$ single sources as of January 22 2024.
The JMDC catalog provides uniform disc angular diameter ($\theta _{\rm UD}$) measurements and the limb-darkened angular diameter ($\theta _{\rm LD}$). In our analysis we used $\theta _{\rm LD}$.

The precision of the calibration of the SBCR depends strongly on the precision of the angular diameter measurements, and we therefore defined quality cuts to keep the best subsample of the JMDC catalog.
The JMDC catalog is a compilation of angular diameter measurements obtained in different studies. These were estimated using different techniques: optical interferometry, lunar occultation or intensity interferometry.
For consistency when comparing angular diameter values, we retained only optical interferometry measurements. 
In addition, we kept angular diameter measurements with a value over error higher than $20$.
In the catalog, $429$ stars ($67$ main sequence stars, see below how we selected giant stars) have more than one stellar diameter measurement. From those, we kept only the stars whose measurements agreed within $3\sigma$. We performed tests with different thresholds for consistency and found $3\sigma$ to be best. We compare the effects of choosing a different rejection threshold in units of $\sigma$ in Section~\ref{subsec:fitchoices}. 
In total we discarded $12$ main sequence stars which had inconsistent measurements to remove the scatter caused by the combination of incompatible measurement methods. 

We cross-matched the JMDC sample with Gaia~DR3, and found that $806$ objects out of $1061$ had Gaia matches. 
To perform the cross-match we used \texttt{astroquery} \citep{Ginsburg2019} to find the Gaia~DR3 (or DR2 in case DR3 was not available) and 2MASS IDs of each star from their Simbad ID\footnote{\url{http://simbad.u-strasbg.fr/simbad/sim-fbasic}}, given in the JMDC catalog.
Then we used the Gaia archive\footnote{\url{https://gea.esac.esa.int/archive/}} to retrieve the information from the Gaia~DR3 catalog (proper motion, parallax and $G$, $G_{\rm RP}$ and $G_{\rm BP}$ magnitudes, etc.) and the 2MASS catalog (JHK magnitudes). 
As the low-mass stars in the JMDC catalog are nearby, and therefore have high proper motions, this technique helped us avoid the mismatches which often occur when doing a positional cross-match.
To understand why some stars do not have a cross-match in Gaia, we estimated $G$-mag from 2MASS magnitudes using the transformation given in \citet{Riello2021}. 
We found that almost all the stars that do not have a Gaia cross-match have an estimated magnitude of $G\leq3$, which is the instrument's brightness limit given by the collaboration\footnote{\url{https://www.cosmos.esa.int/web/gaia/dr3}}.

Taking advantage of the quality flags in the Gaia catalog, we included extra quality cuts to the JMDC sample: 1) Flux over error higher than $10$ for the three bands ($G$, $G_{\rm RP}$ and $G_{\rm BP}$); 2) Parallax over error higher than $10$; 3) Renormalised Unit Weight Error (RUWE) $<1.4$ to remove possible binaries and bad fits, and removed stars indicated in the non-single star flag by Gaia; 4) Removed sources marked as variable in Gaia; 5) Distance $<100$\,pc to minimize the effect of extinction.

\begin{figure}[ht!]
\begin{center}
\includegraphics[width=\linewidth]{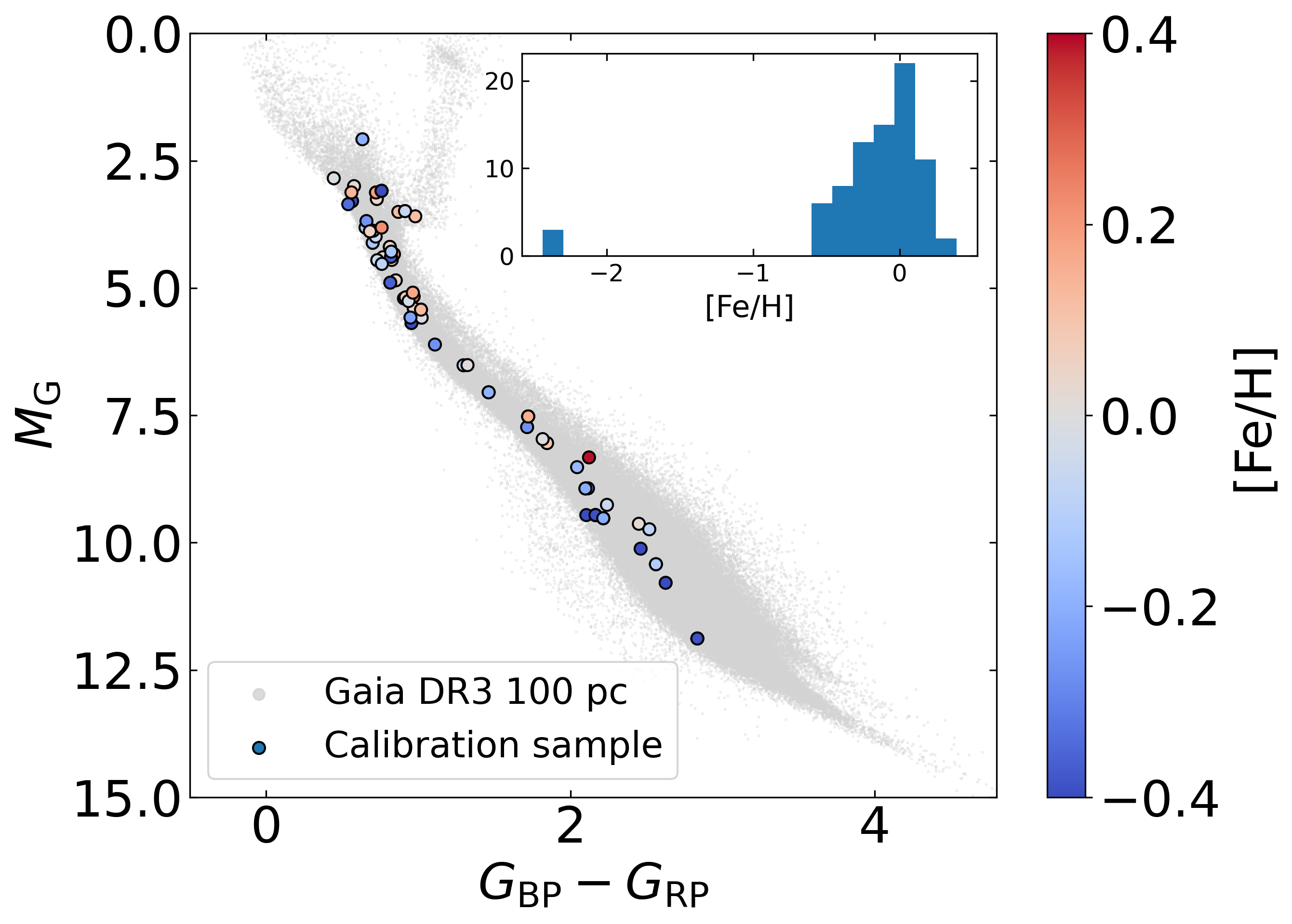}
\caption{Color-magnitude diagram of the JMDC catalog after our quality cuts, color-coded by the literature metallicities [Fe/H]. The distribution of metallicities for our sample is shown in the small panel at the top right of the figure. We also included in light-gray the $100$\,pc sample from Gaia DR3 for comparison. We obtained the $100$\,pc sample from the Gaia archive applying a cut in parallax, together with a value over error $>50$ quality cut for parallax and fluxes in the three Gaia bands.} 
\label{fig:duvert_sample}
\end{center}
\end{figure}

\citet{Salsi2020,Salsi2021} showed that the SBCR is different for main-sequence and giant stars. 
We confirmed this difference using our calibration sample and our giant classification. 
Therefore, for our calibration we kept only main sequence stars.
To separate the giants from the main sequence stars, we approximately selected main sequence stars and we fit a $7^{\rm th}$-degree polynomial, $f(x)$ to the main sequence stars in the color-magnitude diagram ($M_{\rm G}$ versus $(G_{\rm BP}-G_{\rm RP})$ in Figure~\ref{fig:duvert_sample}). The results of the fit was

\begin{equation}
\begin{split}
    f(x) = & 0.07\times x^7 -0.92\times x^6+4.67\times x^5 - \\
    &11.14\times x^4+12.38\times x^3-6.06\times x^2+\\
    &6.27\times x+0.04
\end{split}
\end{equation}

\noindent where $x=G_{\rm BP}-G_{\rm RP}$.
All the stars that were at least 2 magnitudes brighter than $f(x)$ were considered giants. 
In addition, we removed stars that had a spectral type classification of I, II, III or IV in the Simbad database. From the $1061$ single stars in the JMDC catalog, in total we removed $840$ giants.

We limit the fit to the color range $0.42<G_{\rm BP}-G_{\rm RP}<3.2$ where the JMDC catalog provides sufficient data. This cut removed two late type M~dwarfs, and limits our calibration to main sequence stars of spectral type F$0$ to M$4$. 
After all the cuts, we have $70$ main sequence stars, with $85$ angular diameter measurements in total from different studies \citep{Lane2001,Segransan2003,Kervella2004,Kervella2008,Berger2006,Baines2008,Boyajian2008,Boyajian2012a,Boyajian2012b,Boyajian2013,Bazot2011,Bazot2011b,vonBraun2011,vonBraun2012,Crepp2012,Huber2012,Perraut2013,Perraut2015,Perraut2016,White2013,Maestro2013,Howard2014,Mennesson2014,Creevey2015,Jones2015,Kane2015,Kane2017,Tanner2015,Fulton2016,Ligi2016,Ligi2019,Bonnefoy2018,Karovicova2018,Karovicova2020,Schaefer2018,Borgniet2019,Rabus2019,Romanovskaya2019,Wood2019,Salsi2021}.

Metallicity has been shown to have a significant influence on the SBCR \citep[e.g., ][]{Boyajian2012a,Boyajian2012b}. 
Therefore, it is necessary to include a metallicity measurement in the calibration of the SBCR to reduce systematic uncertainties. 
We compiled the metallicity measurements ([Fe/H]) for our sample from the studies mentioned above and other literature measurements \citep{Gaspar2016,Soubiran2016}, and kept in our calibration sample only the $65$ stars with ${\rm [Fe/H]}$ measurements.
Our calibration sample has metallicity values in the range $-2.44 < [{\rm Fe/H}] < 0.39$\,dex, and the metallicity distribution is shown in Figure~\ref{fig:duvert_sample}.

In conclusion, from the $1061$ single stars in the JMDC catalog, after removing all the giant stars ($840$) and applying our quality cuts, our calibration sample contains $65$ main sequence stars with metallicity measurements. This sample will be use in the reminder of the analysis.

\subsection{Fit and results of the Gaia~DR3 SBCR}
\label{subsec:resultscalibration}

\begin{figure}[ht!]
\begin{center}
\includegraphics[width=\linewidth]{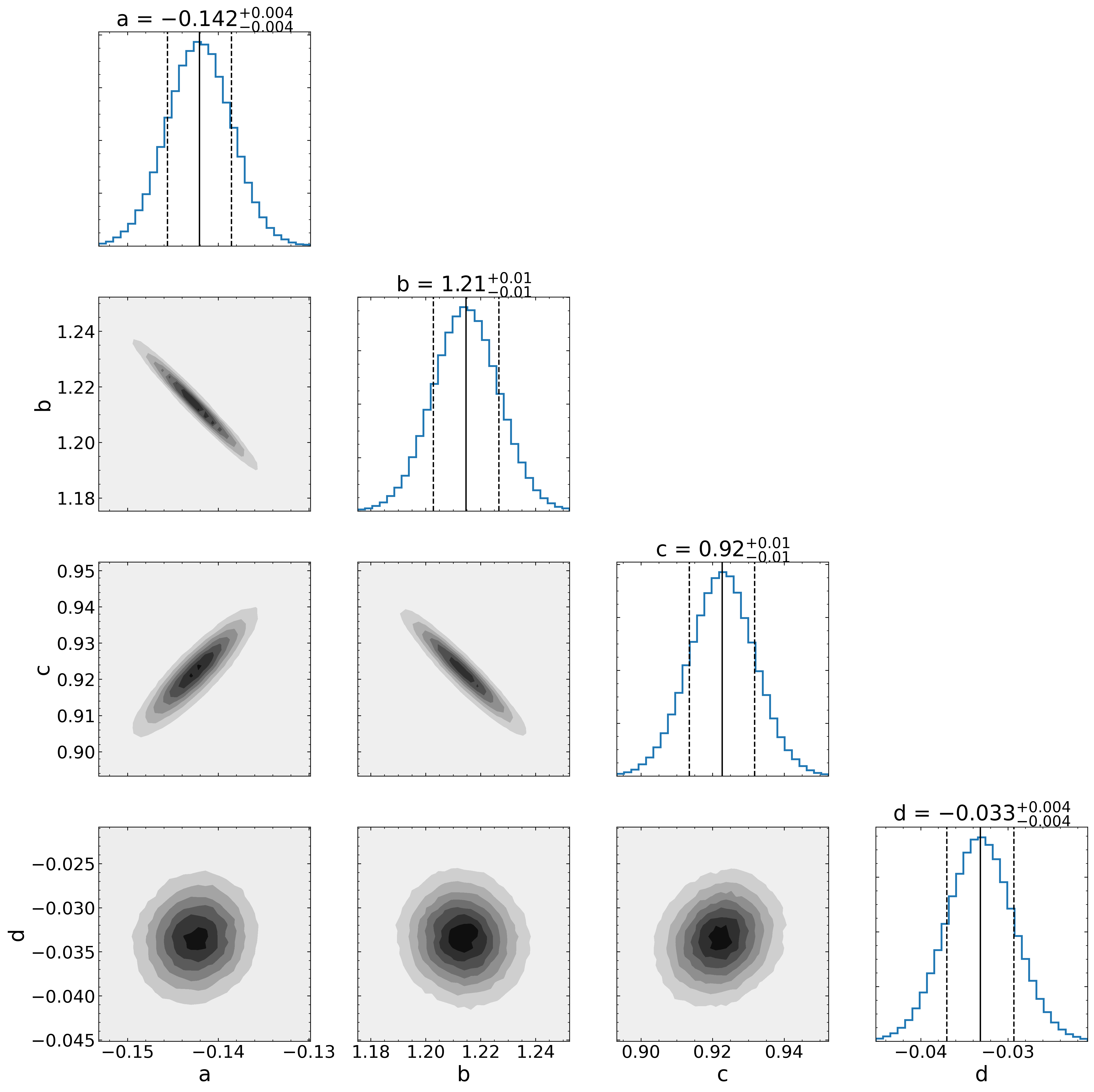}
\caption{Grid plot of the posterior distributions for the four parameters of the likelihood in Equation \eqref{eq:likelihood} ($a$, $b$, $c$ and $d$) obtained with a MCMC-fit using \texttt{emcee} for the case $S_{\rm G}$ and $(G_{\rm BP}-G_{\rm RP})$.} 
\label{fig:grid_plot_g_bp_rp}
\end{center}
\end{figure}

Using our clean subset of the JMDC catalog cross-matched with Gaia~DR3 (Section~\ref{subsec:calibrationsample}), we calibrated the Gaia~DR3 SBCR in Equation~\eqref{eq:SBCR_flux} for main sequence stars.
We used the three Gaia magnitudes ($m_\lambda = G$, $G_{\rm RP}$ and $G_{\rm BP}$), the three colors ($X = G_{\rm BP}-G_{\rm RP}$, $G-G_{\rm RP}$ and $G_{\rm BP}-G$) and a second degree polynomial for the $p(X)$ function in Equation~\eqref{eq:SBCR_flux}. 
To fit this relation we used a Gaussian mixture model. The likelihood for each star is the sum of a Gaussian distribution with the measurement uncertainty and a second, broader Gaussian to account for outliers.  
We modeled the distribution of outliers with standard deviation $\sigma _{\rm out} = 0.3$, which we chose to be larger than the typical measurement uncertainty, and we assumed that the prior probability of a star being an outlier is $1-g$, with $g=0.9$.
We tried different values for $\sigma _{\rm out}$ and $g$ and found that the ones mentioned above were the best suited for our analysis. This test is discussed in Section~\ref{subsec:fitchoices}.
The likelihood of our model is defined as

\begin{align}
        \mathcal{L} = \prod_i \bigg(&A_i \exp\left( \frac{-(\log_{10} S_i - \log_{10} S_{\rm m _{\lambda,i}})^2}{2\sigma_i ^2} \right) \nonumber \\ 
        &+ B_i \exp\left( \frac{-(\log_{10} S_i - \log_{10} S_{\rm m _{\lambda,i}})^2}{2(\sigma_i ^2 + \sigma _{\rm out}^2)} \right) \bigg) \label{eq:likelihood}
\end{align}
with
\begin{align}
A_i&=\frac{g}{\sqrt{2\pi \sigma_i^2}}\\
B_i &= \frac{1-g}{\sqrt{2\pi (\sigma_i^2 + \sigma_{\rm out}^2)}}
\end{align}

\noindent where $\log_{10} S_{\rm m _{\lambda,i}}$ is the quantity defined in Equation~\eqref{eq:SBCR} which we calculated from the angular diameter measurements and Gaia magnitudes, and $\log _{10} S_i$ is calculated from the proposed $p(X)$ such as
\begin{equation}
    \log _{10} S_i = (a\times X^2 + b\times X + c) \times (1 + d\times {\rm [Fe/H]}) \label{eq:model}
\end{equation}

\noindent where $X$ is the color index, and $a$, $b$, $c$ and $d$ are the parameters of the models that we are fitting.
We compared this functional form with a simple polynomial, but found that the one in Equation~\eqref{eq:model} is the most suited for our problem as suggested by \citet{Mann2019}. We discuss the comparison of the functional forms in Section~\ref{subsec:fitchoices}. 
We performed an MCMC-fit using \texttt{emcee} \citep{Foreman-Mackey2013} to sample the posterior composed of the likelihood in Equation~\eqref{eq:model} and flat priors in all four parameters. 
We ensured the convergence of the MCMC by calculating the integrated autocorrelation time ($\tau _f$) and performing $100\tau_f$ steps, which corresponds to ${\sim}100$ independent samples, following the suggestions in the documentation of \texttt{emcee}\footnote{\url{https://emcee.readthedocs.io/en/stable/tutorials/autocorr/}}. 
We performed this fit to obtain a posterior distribution for the parameters $a$, $b$, $c$ and $d$, for the three Gaia magnitudes and the three colors. 
One example of the posterior distribution is shown in the Figure~\ref{fig:grid_plot_g_bp_rp}.
As the distributions of the parameters are nearly Gaussian, we report the uncertainties as the standard deviation in the remainder of the paper. 

\begin{figure*}[ht!]
\begin{center}
\includegraphics[width=\linewidth]{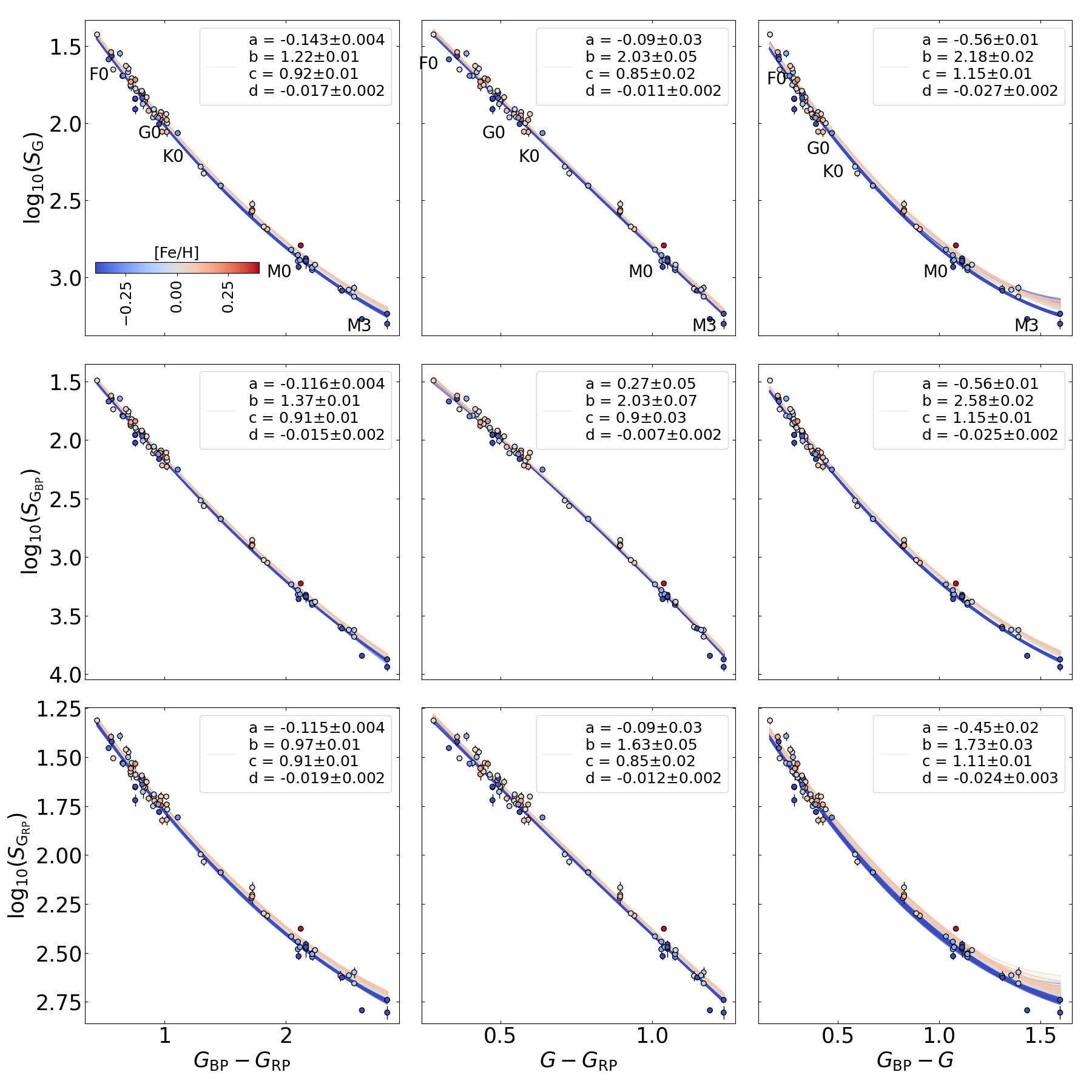}
\caption{Results for our calibration of the Gaia SBCR. We show all the possible combinations between the fluxes (Equation~\eqref{eq:SBCR_flux}) for the three Gaia magnitudes and the three Gaia colors. We color-coded the calibration sample with metallicity, which covers a range from $-2.44<{\rm [Fe/H]<0.39}$\,dex. We included $100$ random samples of the fit for three metallicities color-coded using the same scale as shown in the color-bar: ${\rm [Fe/H]}={-0.5,0,0.1}$. Finally we included the fitted parameters for each case, that are also shown in Table~\ref{table:results_fit}.}
\label{fig:fit_sbcr_gaia}
\end{center}
\end{figure*}

The results of the fit for each combination of Gaia magnitudes and colors are shown in Figure~\ref{fig:fit_sbcr_gaia}, and the values of the coefficients are in Table~\ref{table:results_fit}.
We included $100$ random samples of the fit in Figure~\ref{fig:fit_sbcr_gaia} for three cases of metallicity: [Fe/H]$=\{-0.5,0,0.1\}$\,dex. We show in Section~\ref{subsec:influencemetallicity} that the metallicity correction helps reduce systematic uncertainties when estimating stellar radii. 
More than $80\%$ of the sample is within $2\sigma$ from the fit for all the relations. We also found that the largest offsets and residual scatters are for the lowest-mass stars for the cases where the $(G_{\rm BP}-G)$ color is included. 
This is likely due to the low SNR in the flux on the BP band for the low-mass stars in comparison to the high-mass stars (even though all stars have high SNR $>300$ in this band), combined with the small number of low-mass stars to perform the fit.   
Therefore we suggest choosing between these nine combinations according to the data in each case.
For example, using the fit with $S_{\rm G}$ and $(G-G_{\rm RP})$ to calculate angular diameters of faint M~dwarfs would provide better results in the case where the $G_{\rm BP}$ magnitude has low SNR.


\begin{deluxetable*}{ccccccc}[ht!]
\tabletypesize{\scriptsize}
\tablecaption{Results of the fit for the parameters of the model shown in the likelihood in Equation~\eqref{eq:SBCR_flux}. \label{table:results_fit}}
\tablehead{\colhead{Magnitude} & \colhead{Color} & \colhead{Color range [mag]} & \colhead{$a$} & \colhead{$b$} & \colhead{$c$}& \colhead{$d$}
}\startdata
$S_{\rm G}$&$G_{\rm BP}-G_{\rm RP}$&$0.44-2.83$&$-0.143\pm0.004$&$1.22\pm0.01$&$0.92\pm0.01$&$-0.017\pm0.002$\\ 
$S_{\rm G}$&$G-G_{\rm RP}$&$0.28-1.24$&$-0.09\pm0.03$&$2.03\pm0.05$&$0.85\pm0.02$&$-0.011\pm0.002$\\ 
$S_{\rm G}$&$G_{\rm BP}-G$&$0.17-1.61$&$-0.56\pm0.01$&$2.18\pm0.02$&$1.15\pm0.01$&$-0.027\pm0.002$\\ 
$S_{\rm G_{BP}}$&$G_{\rm BP}-G_{\rm RP}$&$0.44-2.83$&$-0.116\pm0.004$&$1.37\pm0.01$&$0.91\pm0.01$&$-0.015\pm0.002$\\ 
$S_{\rm G_{BP}}$&$G-G_{\rm RP}$&$0.28-1.24$&$0.27\pm0.05$&$2.03\pm0.07$&$0.9\pm0.03$&$-0.007\pm0.002$\\ 
$S_{\rm G_{BP}}$&$G_{\rm BP}-G$&$0.17-1.61$&$-0.56\pm0.01$&$2.58\pm0.02$&$1.15\pm0.01$&$-0.025\pm0.002$\\ 
$S_{\rm G_{RP}}$&$G_{\rm BP}-G_{\rm RP}$&$0.44-2.83$&$-0.115\pm0.004$&$0.97\pm0.01$&$0.91\pm0.01$&$-0.019\pm0.002$\\ 
$S_{\rm G_{RP}}$&$G-G_{\rm RP}$&$0.28-1.24$&$-0.09\pm0.03$&$1.63\pm0.05$&$0.85\pm0.02$&$-0.012\pm0.002$\\ 
$S_{\rm G_{RP}}$&$G_{\rm BP}-G$&$0.17-1.61$&$-0.45\pm0.02$&$1.73\pm0.03$&$1.11\pm0.01$&$-0.024\pm0.003$\\ 
\enddata
\tablecomments{This fit is valid for the metallicity range $-2.44 < [{\rm Fe/H}] < 0.39$\,dex.}
\tablecomments{A Python code and the posterior distribution for each of the parameters of the fit are provided in Zenodo which can be used to estimate radii applying our calibration: \url{https://zenodo.org/records/11401588}.}
\end{deluxetable*}

\subsection{Test fit choices}
\label{subsec:fitchoices}

The fit of the Gaia SBCR to the calibration sample in Section~\ref{subsec:resultscalibration} required several important decisions. Here we discuss each one and the alternatives that we studied before choosing the final fit.

Our calibration sample was described in Section~\ref{subsec:calibrationsample}, and we discussed a $3\sigma$ cut to select consistent measurements of angular diameters between the stars that had more than one measurement. 
After the cross-match with Gaia DR3 and removing the giant stars, we found that $67$ stars had more than one measurement of angular diameter. Out of these stars, $12$ stars had inconsistent measurements, meaning they did not agree within $3\sigma$ ($18\%$ of the stars with repeated measurements). 
We tested this same cut using thresholds of $2\sigma$, $4\sigma$ and $5\sigma$. 
As expected, when the threshold increases, the amount of inconsistent measurements decreases (with cuts of $2\sigma$, $4\sigma$ and $5\sigma$ we found $30\%$, $9\%$ and $7.5\%$ of the stars with repeated measurements were considered inconsistent, respectively).
However we noticed that for the $2\sigma$ sample, the lowest metallicity stars (${\rm [Fe/H]}<-0.5$) were removed. 
We applied our new method to estimate radii for a sample of stars with radius measurements from the literature (\citealt{Mann2015}, see Section~\ref{sec:comparison_literature} for a description of the catalog), and found that when using the fit resulting from the $2\sigma$ sample, our radius estimations had a significant systematic uncertainty compared to the literature generated by metallicity. This systematic uncertainty was removed when using the fit to the $3\sigma$ sample.
Finally, the fit to the $a$, $b$, $c$ and $d$ parameters was similar for the $3\sigma$, $4\sigma$ and $5\sigma$ thresholds, and the MCMC fit with a $3\sigma$ cut took fewer steps to converge than the others.  
We took the more conservative approach to select the best and compatible measurements and chose the $3\sigma$ sample.

We used a mixture model to fit our SBCR, as discussed in Section~\ref{subsec:resultscalibration}. For this fit we assumed the outlier distribution to have a width of $\sigma_{\rm out}$, a larger dispersion than the typical uncertainty value, and each point had a probability of being an outlier of $(1-g)$. 
We performed the fit for the different combinations of Gaia colors and magnitudes for a grid of values for $g=\{0.8,0.85,0.9,0.95\}$ and $\sigma _{\rm out}=\{0.2, 0.25, 0.3, 0.35, 0.4\}$, and compared the results for the four free parameters $a$, $b$, $c$ and $d$.
We found that the best-fit values and confidence intervals on $a$, $b$, $c$, and $d$ parameters do not depend on $\sigma _{\rm out}$ for values $> 0.2$, so we decided to fix $\sigma _{\rm out} = 0.3$. 
In addition, we found that the parameters do not change significantly with $g$. For a fixed $\sigma _{\rm out} = 0.3$, there was a maximum difference in the parameters of less than $1\%$ for the set of values of $g$. Thus we decided to fix $g=0.9$, given that different values did not modify the results. 

Last we discuss the choice of functional form. 
We tried two different functional forms to include $[{\rm Fe/H}]$ in the model: the one indicated at Equation~\eqref{eq:model} and $a\times X^2 + b\times X + c + d\times {\rm [Fe/H]}$.
We found no significant difference between the two in the metallicity range of our calibration sample, however the MCMC-fit for the second case took longer to converge, and in the case of $\log_{10}(S_{\rm G_{RP}})$ with the $(G_{\rm BP} - G)$ color, the fit did not converge according to the criterion described in Section~\ref{subsec:resultscalibration}, even after increasing the total amount of steps.
Therefore we concluded that Equation~\eqref{eq:model} (as confirmed by \citealt{Mann2019}) is more suited to include metallicity in the model.

\subsection{Radius estimation with the Gaia SBCR}
\label{subsec:radius_estimation}

Combining the calibrated Gaia SBCR described in the previous section and the Gaia parallaxes, we can estimate stellar radius. 
For any given star in the valid color and metallicity range, using a Gaia magnitude and a color with the Gaia SBCR we can estimate an angular diameter. 
Then with the star's distance measurement ($D$) from Gaia, we can estimate the stellar radius ($R$) using the following relation where $\theta _{\rm LD}$ is the limb darkened angular diameter:
\begin{equation}
    R = \frac{D\theta _{\rm LD}}{2}
\end{equation}
for $\theta _{\rm LD}$ small, and where $\theta _{\rm LD}$ is in radians\footnote{The $\theta _{\rm LD}$ measurements in \citet{Duvert2016} are published in mas, therefore we converted the units using the equation $\theta _{\rm LD _{rad}} = \theta _{\rm LD _{mas}} \times \pi /(1000\times 180\times 3600)$}.


To study the accuracy of our method, we compared the stellar radii from the JMDC sample calculated using their angular diameters and the parallaxes from Gaia, with the stellar radii calculated using our Gaia SBCR. 
We found that for $68\%$ of the sample, meaning between $16^{\rm th}$ and $84^{\rm th}$ percentiles, the difference between the radii estimated with our method and the original radii is $4\%$ or less. The rest of the sample, has a difference of $10\%$ or less.
This result holds for all the relations shown in Figure~\ref{fig:fit_sbcr_gaia}, meaning it is not dependent on the Gaia band used. 
We note that the JMDC sample contains only bright stars with excellent photometry in the three Gaia bands. For fainter star with larger flux uncertainties, the radius estimation will have larger uncertainties as well.

\subsection{Comparison with SBCR from the literature}

\begin{figure}[ht!]
\begin{center}
\includegraphics[width=\linewidth]{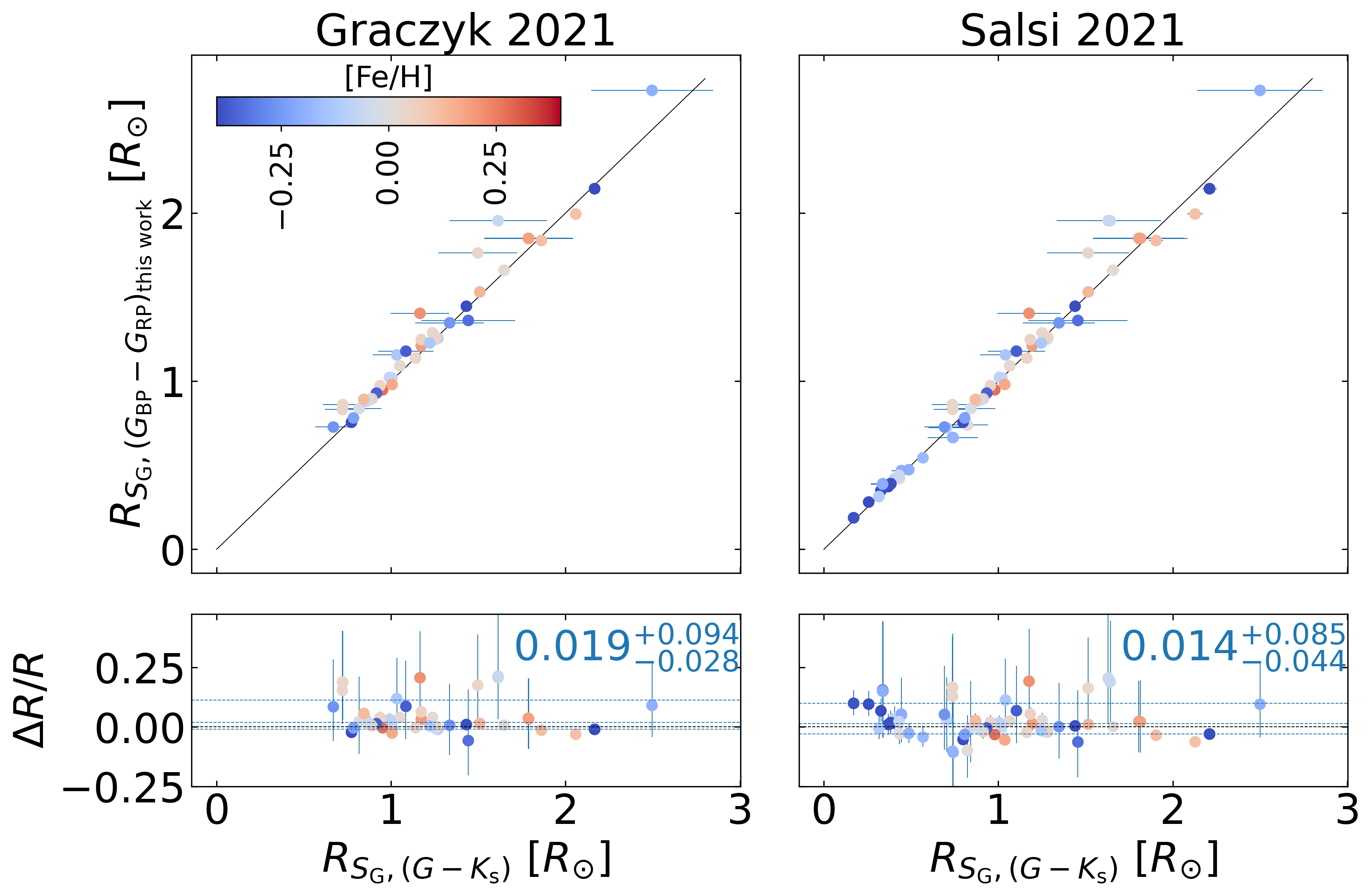}
\caption{Comparison of the results of estimating radii for the JMDC sample using our Gaia SBCR against the results using two recent SBCRs from \citet{Graczyk2021} and \citet{Salsi2021} for the $(G-K_{\rm s})$ color and $S_{\rm G}$. We color-coded the sample with literature metallicities. In the bottom panels we show the fractional radius difference with three horizontal blue dashed lines that show the median, $16^{\rm th}$ and $84^{\rm th}$ percentile, which are also included as blue numbers in each panel.} 
\label{fig:literature_sbcr}
\end{center}
\end{figure}

To facilitate the comparison of our method to literature SBCRs, we used different relations to estimate radii for the clean sample of stars from the JMDC catalog, meaning the stars left after our quality cuts, described in Section~\ref{subsec:calibrationsample}.
Using the most recent SBCRs from \citet{Graczyk2021} and \citet{Salsi2021}, we estimated angular diameters using the Gaia $G$ magnitude and the $(G-K_{\rm s})$ color. 
We then estimated radii from the angular diameter measurements combined with the Gaia parallax, as described in Section~\ref{subsec:radius_estimation}. 
The comparison of the radii estimated using our method to the other two SBCRs is in Figure~\ref{fig:literature_sbcr}. 

Using the \citet{Graczyk2021} SBCR, we found good agreement with our radius estimations. The largest differences corresponds to the objects with large uncertainty in the $K_{\rm s}$ magnitude, which propagates into a large radius uncertainty. For the rest of the stars, the radius measurements agree within $\approx 3\%$. 
Finally, we note that we cannot use the \citet{Graczyk2021} SBCR to estimate the radii of M~dwarfs, given that the relation was not calibrated for such small masses. 

We also found good agreement between our radii and the ones estimated using the SBCR from \citet{Salsi2021}.
We found that the difference between radii is within $\approx 4\%$, except for the stars that have high uncertainty in the $K_{\rm s}$ magnitude, which propagates into a large radius uncertainty. 
In addition, for the low-mass stars ($<0.6~{\rm M_\odot}$), there is a significant systematic where the fractional difference seems to have a negative slope with radius. In Section~\ref{sec:comparison_literature} we compare our radius estimations to two samples from the literature with measured radii (\citet{Mann2015} and the radii from the Exoplanet Archive). Using those same samples, we calculated radii using the SBCR from \citet{Salsi2021} and we compared them to the literature values. We found the same systematic described above in the two samples, therefore we conclude that the SBCR from \citet{Salsi2021} is causing the issue, and thus is not appropriate to estimate radius of low-mass stars.
We also compared the radius estimation using the two literature SBCR, for higher mass stars, and found good agreement with a difference between radii within $1\%$. The main difference in this comparison is that the radius for the stars with high $K_{\rm s}$ uncertainty agree between the two estimations. 

When comparing the uncertainties in the radius estimation we note that our radii can be significantly more precise than in these two studies when the uncertainty of the $K_{\rm s}$ magnitude is large. This shows the advantage of our method to estimate stellar radii: it depends only on Gaia photometry which has higher precision than previous surveys, which translate into smaller radius uncertainties.

\section{Comparison to previous radius measurements}
\label{sec:comparison_literature}

To study the accuracy of our method to estimate stellar radii, we applied our calibration of the Gaia SBCR on several samples from the literature which had radius and metallicity measurements.
The samples we compared to are a set of M~dwarfs from \citet{Mann2015} and a sample of FGK and M dwarfs from the  Planetary Systems Composite Parameters Table from the NASA Exoplanet Archive \citep{NASAEXOP,NASAEXOP2}\footnote{\url{https://exoplanetarchive.ipac.caltech.edu/}}.
The result of the comparisons are shown in Figure~\ref{fig:comparison_lit}.
In addition we compared our radii estimations to the radii from the $50$\,pc sample of main sequence stars from Gaia~DR3, and the results are shown in Figure~\ref{fig:Comparison_gaiadr3}. 
We obtained the $50$\,pc sample from the Gaia archive applying a cut in parallax.
We applied similar quality cuts for the two literature samples as for our calibration sample, described in Section~\ref{subsec:calibrationsample}. 
We required stars to have a distance $<100$\,pc (except for the Gaia~DR3 sample which has a $50$\,pc limit), RUWE $<1.4$, parallax SNR $>10$ and photometric flux in the $G$, $G_{\rm BP}$ and $G_{\rm RP}$ bands with SNR $>10$. 
In addition, we only estimated a radius for stars in the valid metallicity (${\rm [Fe/H]}$) and color range for our calibration described in Section~\ref{subsec:calibrationsample}. 
To estimate radii using our method we used the fit from the Table~\ref{table:results_fit} corresponding to $S_{\rm G}$ and the color $(G_{\rm BP}-G_{\rm RP})$.

\subsection{Comparison to radii from \citet{Mann2015}}

\citet{Mann2015} measured spectra for a sample of M~dwarfs to estimate bolometric flux ($F_{\rm BOL}$) and effective temperature ($T_{\rm eff}$). 
They calculated $F_{\rm BOL}$ by integrating under the radiative flux density and $T_{\rm eff}$ by comparing the spectra to the CFIST suite of the BT-SETTL version of the PHOENIX atmosphere models \citep{Allard2013}. 
\citet{Mann2015} used their measured $F_{\rm BOL}$ and $T_{\rm eff}$ combined with parallaxes from the literature to estimate the radius of each star using the Stefan-Boltzmann law.
They also estimated metallicities using empirical relations between equivalent widths of atomic features like Na and Ca and metallicity, that were calibrated for M~dwarfs using wide binaries with an FGK primary and M~dwarf secondary \citep{Mann2013,Mann2014}.
We cross-matched this sample with Gaia~DR3 to obtain parallaxes and photometry in the Gaia bands, using the same method described in Section~\ref{subsec:calibrationsample}. 
We used the Gaia photometry and parallaxes, together with the $[{\rm Fe/H}]$ measurements from \citet{Mann2015} to estimate radii for the sample with our Gaia SBCR.
In total we compared the radii estimation from \citet{Mann2015} for $100$ stars with our calculations, and the results are in the left panel of Figure~\ref{fig:comparison_lit}. 
We also included a panel with the fractional radius difference as a function of radius. 
We find that most of our radius measurements agree within $1\sigma$ with the radii from \citet{Mann2015}, with a radius fractional difference smaller than $3.6\%$, and there is no significant offset between the measurements. 
In addition, we do not find a correlation with metallicity.
Therefore we conclude that there is no systematic added by our method. 
The agreement is expected given that \citet{Mann2015} used the angular diameter measurements from the JDMC catalog to calibrate their method, which we also used to calibrate our SBCR.


\subsection{Comparison to radii from the NASA Exoplanet Archive}

\begin{figure}[ht!]
\begin{center}
\includegraphics[width=\linewidth]{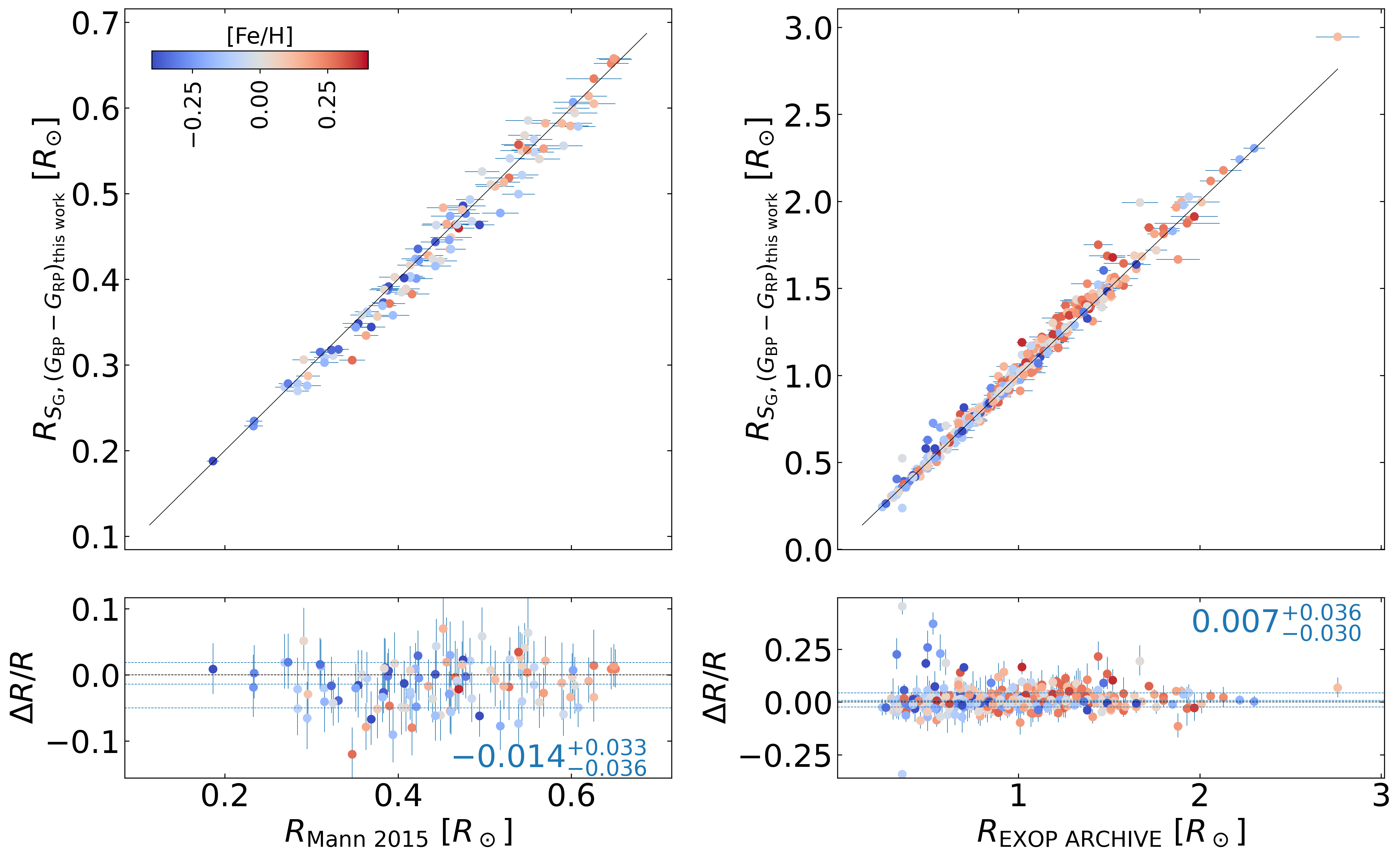}
\caption{We compared the radius estimations using our method with two samples from the literature to asses the accuracy of our method. The two samples are a set of M~dwarfs from \citet{Mann2015} (left panel in the figure) and a sample of FGK and M dwarfs from the Planetary Systems Composite Parameters Table from the NASA Exoplanet Archive (right panel in the figure). We color-coded the stars in each panel by the metallicities provided by each source. We included in the bottom panels the fractional radius difference as a function of the radius in the literature. In these panels we also show the zero difference as a horizontal black line, and in a blue dashed line we show the median, $16^{\rm th}$ and $84^{\rm th}$ percentile, which are also included as blue numbers in the panel. We found good agreement between the measurements.} 
\label{fig:comparison_lit}
\end{center}
\end{figure}

The NASA Exoplanet Archive contains a compilation of all the known exoplanets with their characteristics and the properties of the host star. 
We used the Planetary Systems Composite Parameters Table \citep{NASAEXOP2}, which contains a vetted sample of the archive with the best estimation of parameters.
Among those properties are radius and metallicity. 
In general, the calculations of radius and metallicities are done similarly to the work done by \citet{Mann2015}, where the fitting of observations is done on spectra or spectral energy distributions (SEDs).
In some cases, the [Fe/H] was estimated using fitting of theoretical stellar models.
We kept only the host stars with radii over uncertainty larger than $10$ to remove a few clear outliers with large radius uncertainty. 
We cross-matched the sample of host stars with Gaia~DR3 using the same method described in Section~\ref{subsec:calibrationsample}. 
With the Gaia~DR3 photometry and parallax, and metallicity we estimated radii using our Gaia SBCR. 
In total we used $434$ stars from the NASA Exoplanet Archive to compare to our method, and the results are in the right panel of Figure~\ref{fig:comparison_lit}.
For most of the stars, our measurements and the values in the Exoplanet Archive agree within $1\sigma$, with a radius fractional difference smaller than $3.6\%$, and there is no significant offset.   
For $6$ stars in the sample we found that our radii estimation were larger than the literature value by $\Delta R/R {\sim} 0.5$. All of these radii where estimated by the same study \citep{Demangeon2021}. We found that all $6$ stars had more than one radius estimation in the Archive, and that the second measurement agreed with our estimations. Therefore we decided to remove them from the comparison.
The comparison with the Exoplanet Archive sample has several outliers ($18$ stars with $\Delta R/R \gtrsim 0.1$) which are likely due to the combination of different methods to estimate radius in the Archive. 
However, in total the outliers account for $4\%$ of the sample, and they do not modify significantly the median difference or the scatter.
In addition we found no significant trend with metallicity. 
Therefore we conclude that our method does not add systematic uncertainty to the radius calculations, and it is estimating accurate radii.

\subsection{Comparison to radii from Gaia~DR3.}
\label{subsec:comparisongaia}

As part of Gaia~DR3, the collaboration released stellar radius measurements.
Therefore we decided to compare our estimation of radii using the Gaia SBCR with the results from Gaia.
We used the Gaia Archive to extract photometry, parallaxes, radius measurements and metallicities for the $50$\,pc sample, which we selected applying a parallax cut. The Gaia collaboration used two methods to estimate radius \citep{Andrae2022}.
One is the General Stellar Parameterizer from Photometry (GSP-Phot).
This method utilizes a main algorithm (Aerneas \citealt{Bailer-Jones2011}) which fits the measured BP/RP spectra \citep{DeAngeli2022}, parallax and G magnitude to estimate stellar age, mass, metallicity ([M/H]) and extinction using stellar PARSEC iscochrones (PARSEC 1.2S Colibri S37 models \citealt{Tang2014,Chen2015,Pastorelli2020}),
and four different stellar atmosphere models: MARCS for $2,500<T_{\rm eff}<8,000$\,K, PHOENIX for $3,000<T_{\rm eff}<10,000$\,K, A-stars models for $6,000<T_{\rm eff}<20,000$\,K, and OB stars models for $15,000<T_{\rm eff}<55,000$\,K \citep{Creevey2022}.
The Gaia collaboration also uses the General Stellar Parametrizer from Spectroscopy (GSP-Spec \citealt{Recio-Blanco2022}) method to estimate stellar atmospheric parameters and individual chemical abundances ($T_{\rm eff}$, $\log g$, [M/H], [Fe/H], etc) from the Radial Velocity Spectrometer (RVS) spectra of single stars, with a resolution of $R{\sim}11500$.
The second method used by the Gaia collaboration to estimate radii is the Final Luminosity Age Mass Estimator (FLAME). FLAME uses BaSTI stellar evolution models \citep{Hidalgo2018} assuming solar metallicity to estimate mass, age and evolutionary stage, and it also produces radius, luminosity and gravitational redshift. These estimations are done based on $T_{\rm eff}$, $\log g$ and [M/H] from GSP-Phot or GSP-Spec. 

\begin{figure}[ht!]
\begin{center}
\includegraphics[width=\linewidth]{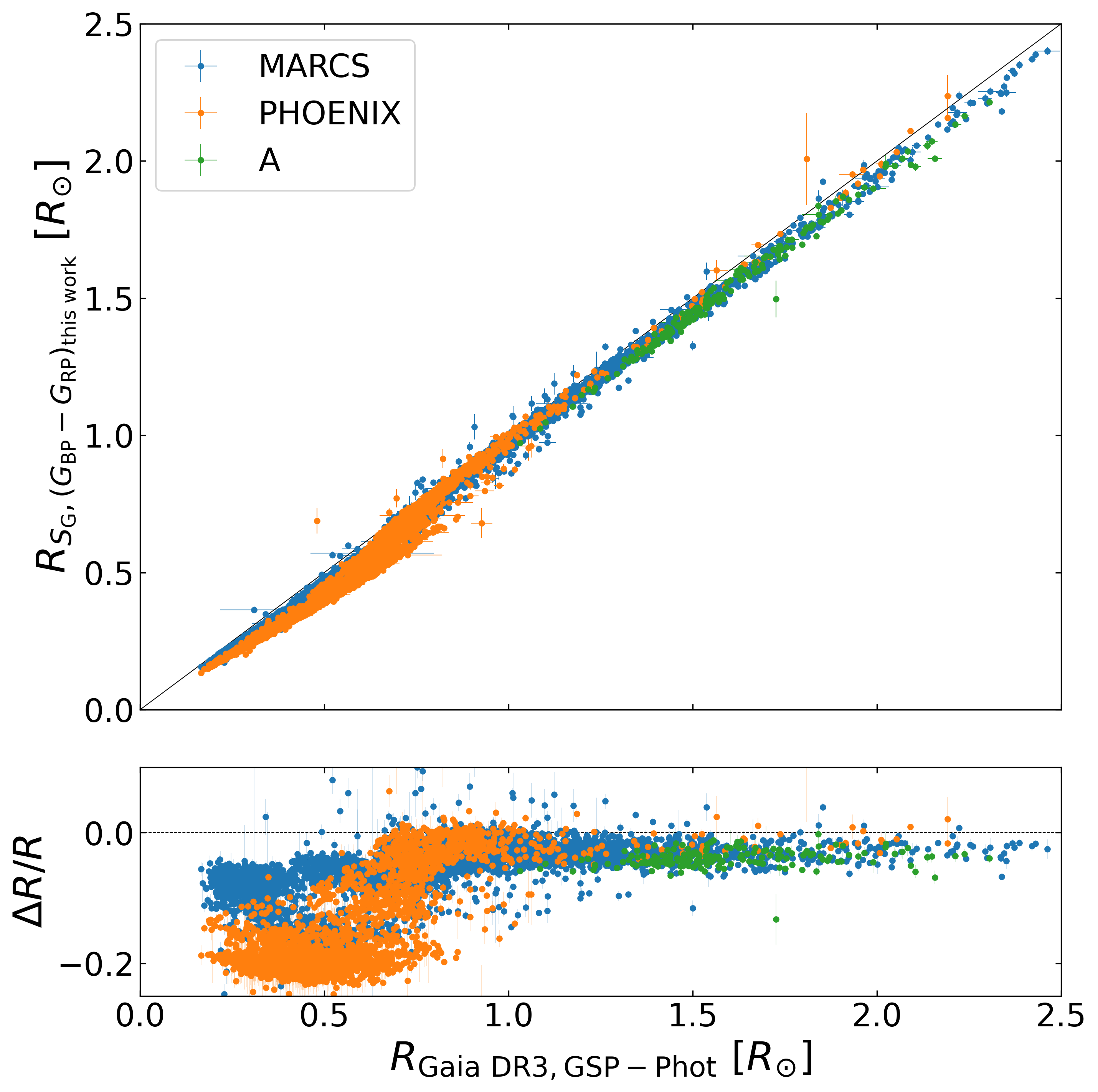}
\caption{Comparison of the radius measurements using our Gaia SBCR and the radii in the Gaia~DR3 catalog for the $50$\,pc sample. We show in this figure only the radii calculated by Gaia with the GSP-Phot method, and we color-coded the stars by the model used in this calculation. In general we found that our radius estimations are systematically smaller than the values in the Gaia catalog, and that the shift is different according to the model used and the radius. We found that for radius $>0.6\,{\rm R_\odot}$, the systematic difference is $3-4\%$ for the MARCS and A-stars models, but the difference is slightly smaller ($2\%$) using the PHOENIX models. For the M~dwarfs (radius $<0.6\,{\rm R_\odot}$) the offset is between $10-25\%$ for GSP-Phot using PHOENIX, $7\%$ for the MARCS models and $13\%$ for the FLAME method using GSP-Phot parameters.} 
\label{fig:Comparison_gaiadr3}
\end{center}
\end{figure}

We estimated radii for the $50$\,pc sample of main sequence stars from Gaia~DR3 using our Gaia SBCR and Gaia photometry, parallaxes, and assuming solar metallicity.
The comparison of our results and the radius estimation using the GSP-Phot method are shown in Figure~\ref{fig:Comparison_gaiadr3}. 
We also compared our radii estimates to the values from GSP-Spec and Flame, but we did not include a figure. 
In general we found that the radii estimated with our method are systematically smaller than the values in the Gaia catalog, but the difference depends on the models used and the radius.
We found that for a radius $>0.6\,{\rm R_\odot}$, our radius estimations are $3-4\%$ smaller than the values in the Gaia catalog calculated using the MARCS or A-stars models, and a slightly smaller shift ($2\%$) using the PHOENIX models.  
In the case where the radii were estimated with the FLAME method with parameters from GSP-Phot, we found also a $4\%$ offset for radius $>0.6\,{\rm R_\odot}$, and $<2\%$ for FLAME using parameters from GSP-SPEC. 
For the M~dwarfs (radius $<0.6\,{\rm R_\odot}$) the offset is between $10-25\%$ when the radii from Gaia were estimated using the GSP-Phot method with the PHOENIX model and $7\%$ for the MARCS models. We found a difference of $13\%$ when the Gaia radii were estimated using the FLAME method with parameters from GSP-Phot. 
The difference between the MARCS and PHOENIX models was already discussed by \citet{Andrae2022}.
We conclude that the radii estimations from Gaia should be used with care. We found that the best agreement between our calculations and the results in the Gaia catalog for higher mass stars ($>0.6\,\Rsun$) are for the FLAME method using the GSP-SPEC models and for the low-mass stars ($<0.6\,\Rsun$) the best agreement is using the GSP-Phot method with the MARCS models.
However, for low-mass stars these radii calculations have an offset ($7\%$) as well, which should be taken into account. 

To study this offset between radius estimations in more detail, we used our calibration sample of main sequence stars with angular diameter measurements, described in Section~\ref{subsec:calibrationsample}. 
We used the angular diameters and the parallaxes from Gaia~DR3 to estimate radii for the sample and we compared these with the Gaia radius estimations using GSP-Phot. 
We found the same $3\%$ offset between the two radius estimations for radius $>0.6\,{\rm R_\odot}$, and around $10\%$ for smaller stars. 
Therefore we conclude that the offset is not being generated by our method. 
We note that \citet{Fouesneau2022} performed a similar comparison for the Gaia radius estimations using the GSP-Phot method against angular diameters measurements, but included giants in the analysis that make the difference less evident.

The difference between our calculation and the Gaia radii for M~dwarfs is not surprising given that evolutionary and atmospheric models have been shown to still have several discrepancies with observed data \citep[e.g., ][]{Baraffe2015,Dieterich2021}.
Significantly, the models used in the Gaia catalog for M~dwarfs over-predict the radius for all these stars, and not just the magnetically active.


\section{Possible biases}
\label{sec:biases}

To further test the precision of our method of using the Gaia SBCR to estimate radius, we study different properties that could affect our measurements, given that we rely on photometry and parallax.
In this section we discuss the following properties: metallicity, variability and extinction.

\subsection{Influence of metallicity}
\label{subsec:influencemetallicity}

Prior studies estimated radii using stellar models and compared measured radii to models which showed the importance of having precise metallicity values \citep{Burrows2011,Irwin2011b,Feiden2012a,Spada2012,Morrell2019}. Other studies showed that stars with different metallicities follow a different SBCR \citep{Boyajian2012a,Boyajian2012b}.
Therefore we calibrated our Gaia SBCR taking into account the metallicity ([Fe/H]) of our sample. 
However, it is common in the literature to assume solar metallicity ([Fe/H]\,$=0$) when a measurement is not available.
In order to test this assumption applying our method, we used the literature samples discussed in Section~\ref{sec:comparison_literature} to compare the radius estimations that account for metallicity with those that assume solar metallicity. 

In Section~\ref{sec:comparison_literature} we used the offset and scatter between our radius estimation and the literature to study the precision of our method. To study the effect of metallicity, we assumed solar metallicity for all the stars and re-calculated the offset and scatter. For the samples from \citet{Mann2015} and the Exoplanet Archive which had metallicity measurements, we did not find a significant change in the offset or the scatter of the stars. This is not surprising given that the stars in these samples are close to solar metallicity (the bulk of the stars are in the range $-0.3<[{\rm Fe/H}]<0.3$\,dex). 


In addition, when we assumed solar metallicity to estimate radii with our method, we found that for the \citet{Mann2015} and the Exoplanet Archive samples we underestimated radii for low-metallicity stars and overestimated radii for high-metallicity stars.
We found that for stars with $-1.5<{\rm [Fe/H]}<0$\,dex assuming solar metallicity underestimates the radius by about $3\%$, and for stars with $0<{\rm [Fe/H]}<0.4$\,dex assuming solar metallicity overestimates the radius by $2\%$. 
Outside of this range, assuming solar metallicity generates a larger loss of precision. 
In this work we study radius inflation, which is normally a $7-10\%$ difference in radius. Therefore we chose to work with stars with metallicities between $-0.3<{\rm [Fe/H]}<0.3$\,dex for the rest of our analysis, to minimize the radius offset due to metallicity.

\subsection{Influence of variability}

Measurement error due to variability could also affect our calibration of the SBCR and the radius estimation using our method, which depends strongly on photometry. 
Magnetic activity can generate short and long term variability in the photometry of a star \citep[e.g.,][]{Irving2023}. 
It has been shown that measurements in eclipsing binaries are affected by variability \citep{Morales2010,Han2019} or the presence of large polar spots \citep{Morales2010,Kraus2011,Torres2019}. 
To test the effect of photometric variability on the radius estimation, we used the light-curves provided by Gaia~DR3.
We used the sample of M~dwarfs with known ages cross-matched with Gaia DR2 from \citet{Kiman2021}, and we used the Gaia DR2 id to obtain the Gaia~DR3 id.
We selected the stars from this sample that were indicated to be members of the Pleiades cluster. This young cluster is $120$\,Myr old \citep{Dahm2015}, with ${\rm [Fe/H]} = +0.03 \pm 0.05$\,dex \citep{Soderblom2009} and its members are known to show high variability \citep[e.g., ][]{Kiman2021}.
We note that this test is a proof of concept of the effect of high variability when calculating radii. However, our method to estimate radii using the Gaia SBCR is not calibrated for pre-main sequence stars.

The acquisition of the Gaia light-curve works as follows: first Gaia takes a photometry measurement in the G band, then $\approx27$ minutes later a measurement in the BP band and $\approx7.5$ minutes later a measurement in the RP band. 
The time between these sequences of three band measurements varies. 
However, this time is always larger than the time between band observations.
In addition, most of the stars for which we downloaded a light-curve, were observed for $900$ days in total, but not continuously. 
In conclusion, we assumed that the photometry points in each band were taken at the same time. 
Therefore, we can combine the light-curves in each band with the parallax and estimate a time sequence of radius using our method. 

\begin{figure}[ht!]
\begin{center}
\includegraphics[width=\linewidth]{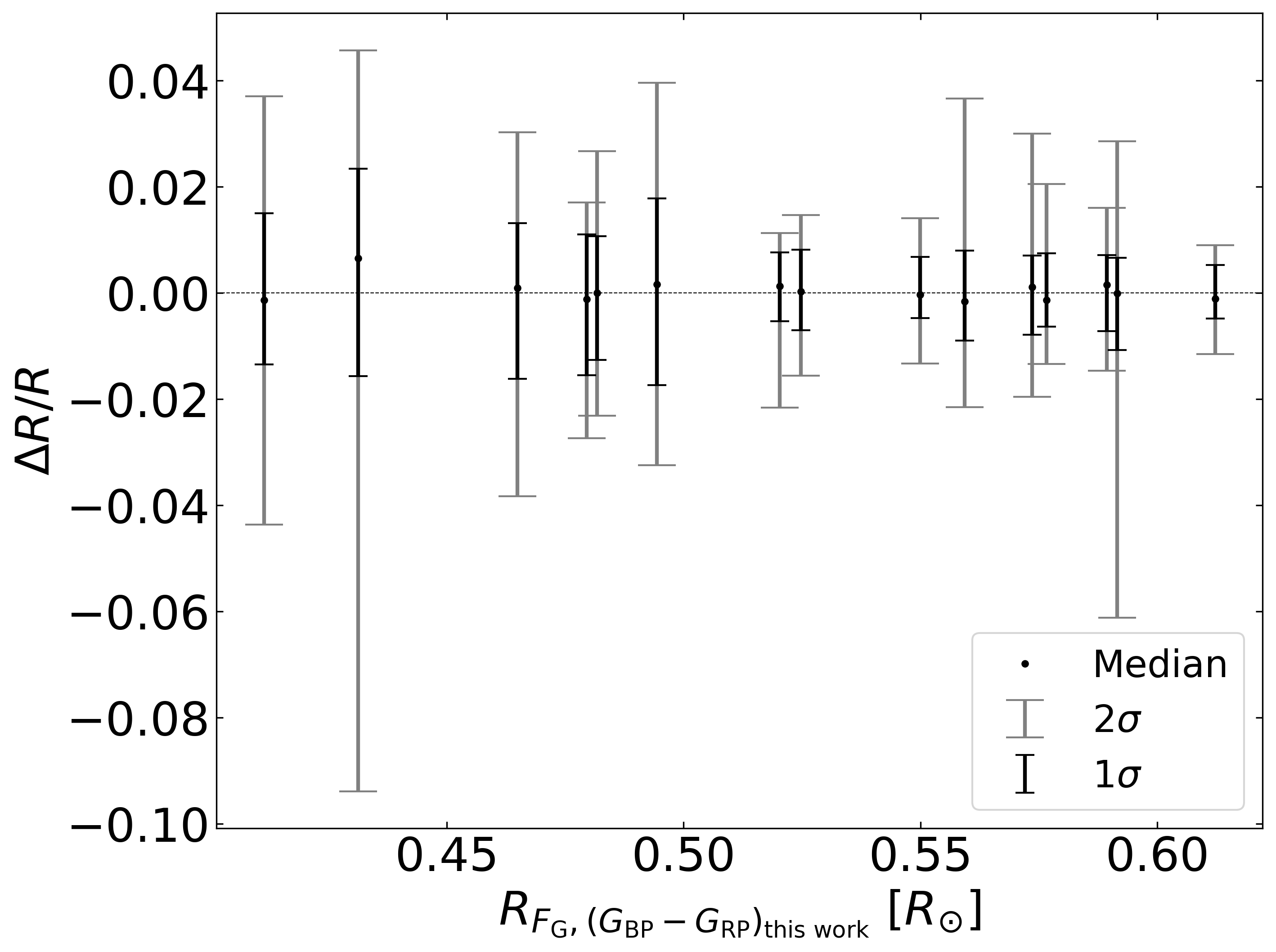}
\caption{Fractional radius difference between our radius estimation using the published Gaia photometry and the median of all the photometric points taken by Gaia for each star. We included the $1\sigma$ and $2\sigma$ deviations to show what the radius difference could be if Gaia would take only one photometry point per star. These stars are from the Pleiades cluster ($120$\,Myr) so we expect high photometric variability. Given that Gaia averages several photometry points to estimate the photometry of the star, photometric variability it is not an issue when estimating radii using Gaia photometry.} 
\label{fig:ple_variability}
\end{center}
\end{figure}

We used this time sequence of radii to estimate the fractional difference between these radii and the radius calculated with the published Gaia photometry for each star.
The median of the fractional difference for each object in our sample as a function of the radius estimated from the published photometry are in Figure~\ref{fig:ple_variability}. 
These stars are young and variable in their photometry due to phenomena such as spots and magnetic cycles \citep{Irving2023}. 
Our analysis shows that if only one photometry point is taken for these stars, the change in radius can be as large as $4\%$ for most of the stars and even $7\%$ and $10\%$ for two stars in the sample.
We note that all the stars in our sample are M~dwarfs, therefore these are fast rotating stars.
However, given that Gaia takes several photometry measurements and then averages all of them for the final photometry, the effect of variability is reduced, as shown by the median difference in radius being $<1\%$ for all the stars in Figure~\ref{fig:ple_variability}. 
Therefore, when using Gaia photometry variability is not a concern for calculating the radius of stars, or for the calibration of the surface brightness-color relation.  

\subsection{Influence of extinction}

In this work we used only nearby stars for most of our analysis ($<100$\,pc) to minimize the effect of extinction on the photometry. 
In reality, most stars are not closer than $100$\,pc, and extinction will affect their photometry. 
An extinction correction is necessary in these cases to use the method described in this paper. 
To approximate the effect extinction would have on radius measurements, we selected a random sample of $20,000$ stars within $500$\,pc from the sample of \citet{Anders2022}. In this work, the authors used the code $\texttt{StarHorse}$ to estimate distances, extinctions (at $\lambda = 542$ nm), ages, masses, effective temperatures, metallicities, and surface gravities for field stars, from the parallaxes and photometry from Gaia EDR3 \citep{Gaia2021}, combined with photometry from 2MASS \citep{Cutri2003}, AllWISE \citep{Cutri2013}, Pan-STARRS1 \citep{Chambers2016}, and SkyMapper \citep{Onken2019}.   

To clean the sample we removed giant stars according to their position in the color-magnitude diagram, using the same method described in Section~\ref{subsec:calibrationsample}. 
We estimated radii for the resulting sample using our method with the corrected and uncorrected magnitudes, and calculated the percentage of radius difference between these two values. 
We repeated this process for all nine relations we derived for the three Gaia colors combined with the three magnitudes, which are shown in Table~\ref{table:results_fit}.
In this random subsample from \citet{Anders2022}, we found that the effect of extinction on magnitudes increases with distance and it is larger on the $(G_{\rm BP}-G_{\rm RP})$ color, as expected. However, the effect of extinction on the radius estimation is more complex, and it is described below.

We found that the extinction correction can generate a radius difference of up to $10\%$, for stars within $500$\,pc. In addition, the influence of extinction on the radius estimation does not depend on which magnitude is used to estimate radius. 
As expected, we found that on average the influence of extinction in the radius estimation increases as a function of the distance, however the increase of the absolute value of the radius difference as a function of distance depends on the color used to estimate the radius and the spectral type of the star. 
To study this effect we divided the sample in three bins of spectral type: F and G dwarfs, K dwarfs and M dwarfs. We found that the influence of extinction for F,G and K dwarfs is similar. For these three spectral types we found that the radius estimated with any of the three Gaia color is larger when the magnitudes are not corrected for extinction. The extinction effect is larger when the $(G-G_{\rm RP})$ color is used to calculate the radius ($6\%$ average radius difference at $500$ pc), while the radius estimation is almost not modified ($0.1\%$ at $500$ pc) when the $(G_{\rm BP}-G)$ color is used.
For M dwarfs we found that when using the $(G-G_{\rm RP})$ color to estimate radii, the radii estimated with magnitudes without the extinction correction are larger, but using the $(G_{\rm BP}-G_{\rm RP})$ and $(G_{\rm BP}-G)$ colors, the radii calculated with magnitudes without extinction correction are smaller. 
Finally, we found that using the $(G_{\rm BP}-G)$ to estimate radii, the average radius difference is of $6\%$ at $500$\,pc, the largest between the three Gaia colors. Using $(G-G_{\rm RP})$ and $(G_{\rm BP}-G_{\rm RP})$ we found an average radius difference of $2\%$ when the $G$ or $G_{\rm RP}$ magnitudes are used. When the $G_{\rm BP}$ magnitude is used in combination with the $(G-G_{\rm RP})$, we found an average radius difference of $5\%$.
In conclusion, the radii estimated using the $(G-G_{\rm RP})$ color are the most affected by extinction for F, G and K, while for M dwarfs $(G_{\rm BP}-G)$ is the most problematic color.

\section{Radius inflation in M~dwarfs}
\label{sec:radiusinflation}

We used the Gaia~DR3 SBCR to study the relationship between radius and magnetic activity and mass for low-mass stars. 
To do this, we measured radius and mass for the \citet{Kiman2019} M~dwarf sample which has magnetic activity measurements and estimates of metallicity.

\subsection{The \citet{Kiman2019} sample}
\label{subsec:kiman2019sample}

\citet{Kiman2019} compiled a sample of M~dwarfs which have $\halpha$ equivalent width ($\haew$) measurements from spectra from the Sloan Digital Sky Survey \citep[SDSS, ][]{West2011,Schmidt2015}, which is a magnetic activity indicator \citep[e.g., ][]{West2011,Newton2017,Kiman2019,Kiman2021}. This sample also has SDSS and 2MASS photometry.
We used the Gaia DR2 source ids provided in the catalog to obtain the corresponding Gaia~DR3 source from the cross-match available in the Gaia archive. We refer to \citet{Kiman2019} for details on the Gaia cross-match.

We applied similar quality cuts as described in previous sections to estimate precise radii (see Sections~\ref{subsec:calibrationsample} and \ref{sec:comparison_literature}). 
We kept stars with parallax measurements with SNR\,$>40$ (not $>10$ as in previous section because these stars are at a larger distance: $100-200$\,pc), photometry in the $G$ and $G_{\rm RP}$ Gaia bands with flux with SNR\,$>10$ and RUWE\,$<1.4$.
Following \citet{Kiman2019}, we selected stars that do not need an extinction correction using the estimation of extinction from the $(r-z)$ color available in the catalog.
We estimated stellar mass using the relation between mass and absolute 2MASS $K_{\rm s}$ magnitude from \citet{Mann2019}. 
In order to estimate precise masses, we kept stars with $K_{\rm s}$ uncertainty $<0.07$\,mag ($\approx$7\% uncertainty in flux). 
After the quality cuts we were left with $7,640$ stars. 
The stars from \citet{Kiman2019} have measurements of $\zeta _{\rm TiO/CaH}$ calculated by \citet{West2011}. 
$\zeta _{\rm TiO/CaH}$ is a metallicity dependent parameter for M~dwarfs defined by \citet{Lepine2007} as the relation between the CaH and TiO molecular indices such that 
\begin{equation}
    \zeta _{\rm TiO/CaH} = \frac{1-{\rm TiO5}}{1-{\rm [TiO5]_{Z_\odot}}}
\end{equation}
where
\begin{equation}
    \begin{split}
        {\rm [TiO5]_{Z_\odot}} = &-0.164\times ({\rm CaH}2+{\rm CaH}3)^3\\
        &+0.670\times ({\rm CaH}2+{\rm CaH}3)^2\\
        &-0.118\times ({\rm CaH}2+{\rm CaH}3) - 0.050.
    \end{split}
\end{equation}

\citet{Woolf2009} calibrated the relation between $\zeta _{\rm TiO/CaH}$ and [Fe/H] by measuring the two quantities for a sample M~dwarfs with high resolution spectra. 
They fit a linear function to the data such that ${\rm [Fe/H]}=a+b\times \zeta _{\rm TiO/CaH}$ where $a=-1.685\pm 0.079$ and $b=1.632\pm 0.096$.
This relation is valid for $-2<{\rm[Fe/H]}<0.5\,{\rm dex}$ according to Figure 6 from \citet{Woolf2009}. 
We used this calibration to estimate approximate [Fe/H] for our sample. 

\subsection{Mass-radius relation for active and inactive stars}
\label{subsec:radiusinfkiman2019}

\begin{figure}[ht!]
\begin{center}
\includegraphics[width=\linewidth]{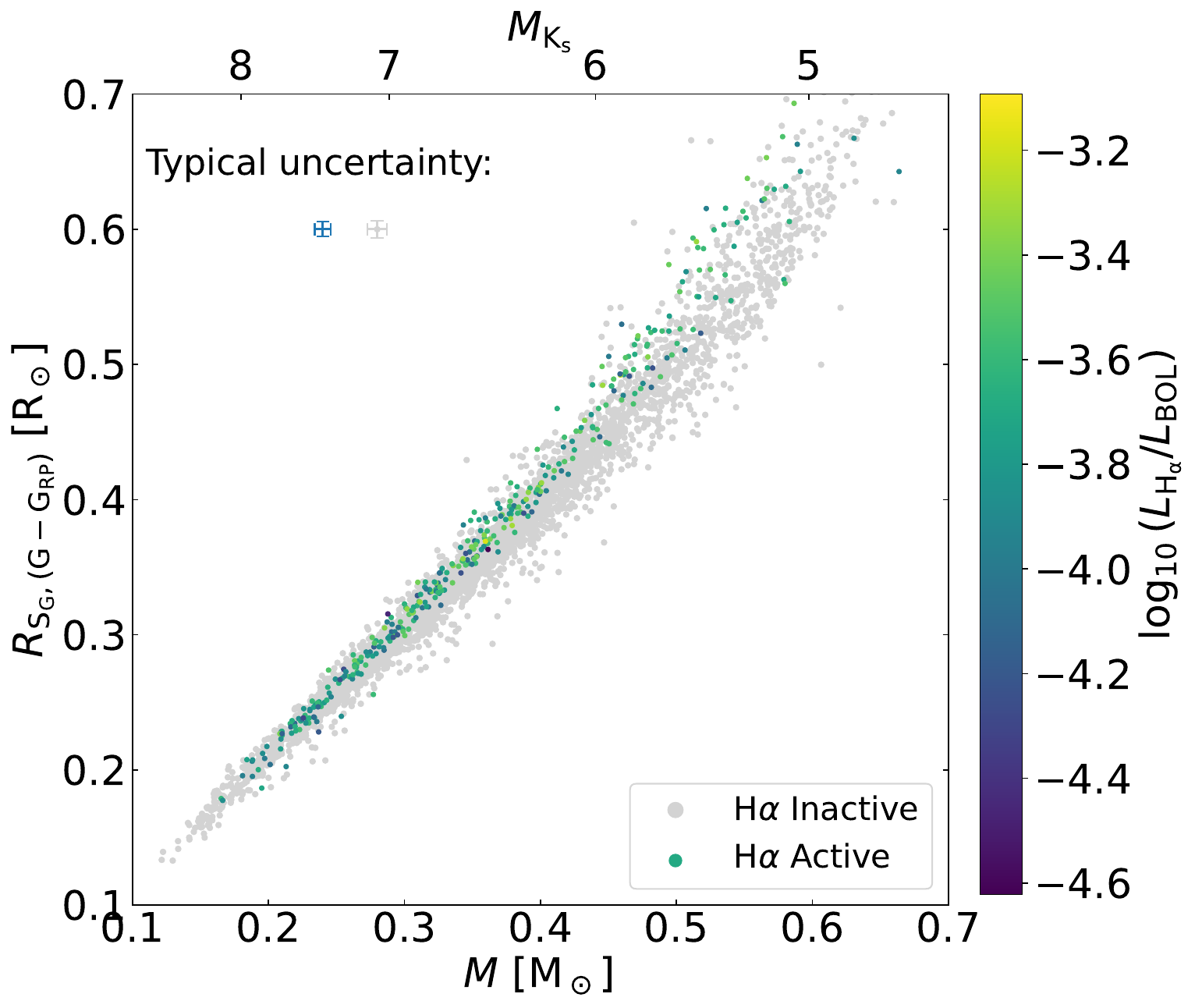}
\caption{Mass-Radius relation for the sample from \citet{Kiman2019}. Inactive stars according to $\halpha$ are shown in gray and active stars are color-coded by fractional $\halpha$ luminosity. We only include the typical uncertainly (median) of the sample. Active stars show larger radii than inactive stars indicating that radius inflation is correlated with magnetic activity.} 
\label{fig:mlsdss_mass_radius}
\end{center}
\end{figure}

We calculated radii for the \citet{Kiman2019} sample using our Gaia SBCR combining the Gaia photometry and parallaxes with the estimated [Fe/H]. 
We used the relation for the $(G-G_{\rm RP})$ color and $S_{\rm G}$ in Table~\ref{table:results_fit}. 
We decided to use this relation instead of the one corresponding to the $(G_{\rm BP}-G_{\rm RP})$ color as in previous sections (see for example Section~\ref{sec:comparison_literature}) because this sample is on average fainter than other samples in this work, and therefore has high uncertainties in the $G_{\rm BP}$ photometry \citep{Kiman2019}.
In addition, we estimated stellar mass using the relation between mass and absolute 2MASS $K_{\rm s}$ magnitude from \citet{Mann2019}, as we discussed in Section~\ref{subsec:kiman2019sample}. 
The relation between our estimated radii and masses are shown in Figure~\ref{fig:mlsdss_mass_radius}. 
We only included the median uncertainty to simplify the visualization of the data. 
The relation from \citet{Mann2019} is valid for masses $<0.7\,{\rm M_\odot}$, so we removed all the stars which had estimated masses outside this range. 
We divided the sample into active and inactive stars with the active flag in the \citet{Kiman2019} sample.
Moreover, we color-coded the active stars according to $\halpha$ by fractional $\halpha$ luminosity ($\lhalbol$), and we show the inactive stars in gray. 
Analysing Figure~\ref{fig:mlsdss_mass_radius}, we found that active stars have larger radii on average than inactive stars all along the M~dwarf regime. 
For this same sample, \citet{Kiman2019} found that active stars are redder in the Gaia color-magnitude diagram than inactive stars (See Figure 17 in their work). Therefore we can conclude that the offset between active and inactive stars is due to an inflated radius rather than underestimated masses.
As $\halpha$ is an indicator of magnetic activity, this result indicates that stars that are magnetically active have, on average, larger radii than their inactive counterparts with the same mass, which indicates that radius inflation is correlated with magnetic activity. 
In addition, the difference in radius between active and inactive becomes smaller for smaller stars, which agrees with previous results \citep[e.g., ][]{Lopez-Morales2007,Morales2022}.

The result in Figure~\ref{fig:mlsdss_mass_radius} agrees with the results from \citet{Stassun2012}. In that work the authors estimated radii from $T_{\rm eff}$ and bolometric luminosity measurements, and they approximated $T_{\rm eff}$ from spectral type, and bolometric luminosity from 2MASS $K_{\rm s}$ magnitudes and parallaxes. They used the isochrones from \citet{Baraffe1998} to estimate masses from the bolometric luminosity. 
In comparison, our parameters are fully empirical given that we rely only on measured data, and our radius estimations were done directly from a SBCR, instead of depending on spectral type.
In addition, we have a larger sample which allowed us to study the difference between active and inactive stars as a function of mass.

\begin{figure}[ht!]
\begin{center}
\includegraphics[width=\linewidth]{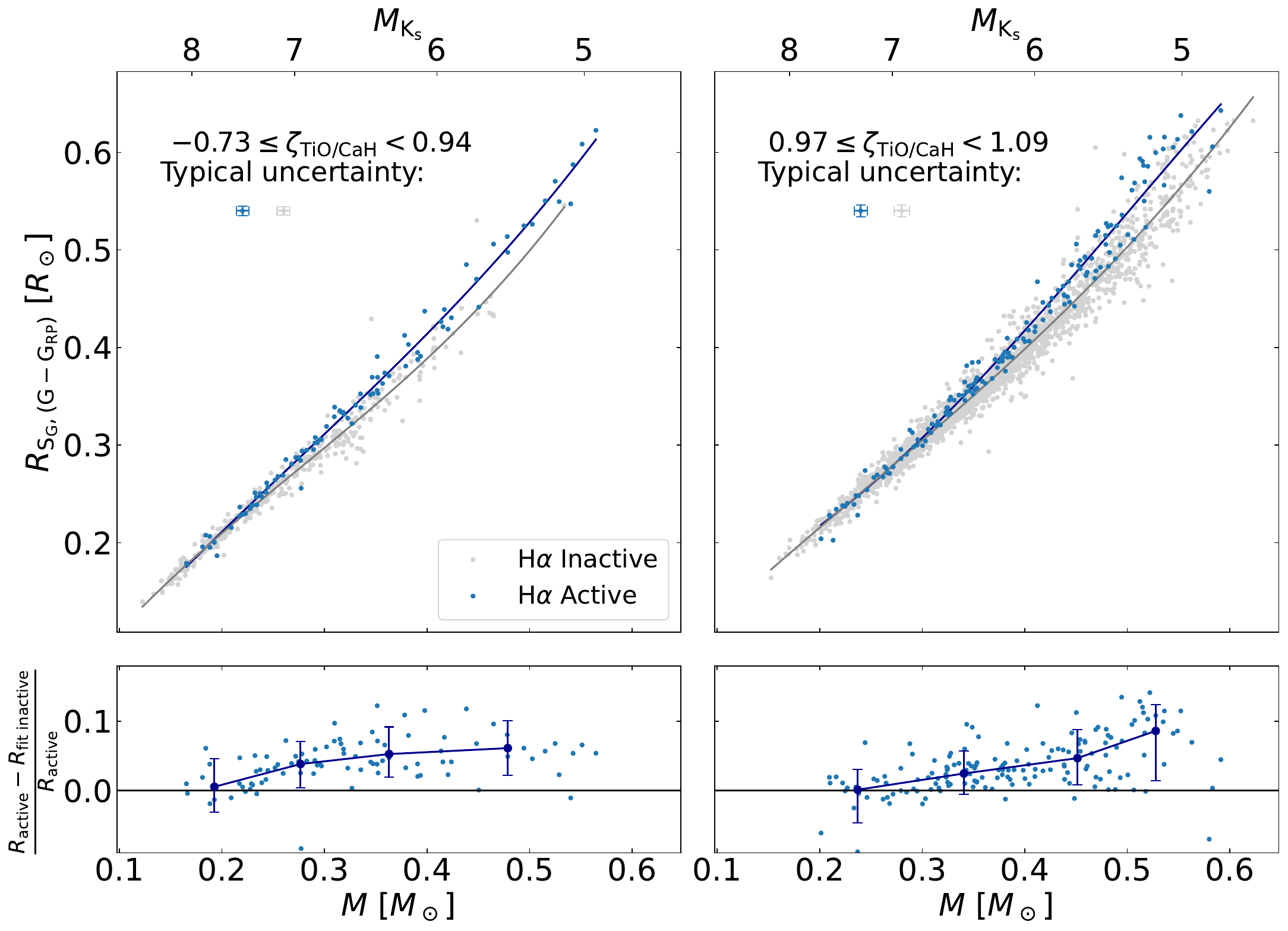}
\caption{Same data as in Figure~\ref{fig:mlsdss_mass_radius} divided in two bins of metallicity relation with metallicity: $0.73\leq\zeta _{\rm TiO/CaH}<0.94$ (left panel) which corresponds roughly to $-0.5\leq{\rm [Fe/H]} < -0.15$\,dex and a solar metallicity bin of $0.97\leq \zeta _{\rm TiO/CaH}<1.09$ (right panel) which is approximately $-0.1 \leq {\rm [Fe/H]} < 0.1$\,dex. We include in the lower panels the fractional radius difference between the radii of the active stars and the fit to the inactive stars for the same mass. In dark blue we show the median and standard deviation for four bins in mass. When we separate the stars by metallicity, the overlap between active and inactive stars seen in Figure~\ref{fig:mlsdss_mass_radius} decreases.} 
\label{fig:mlsdss_mass_radius_feh}
\end{center}
\end{figure}

Figure~\ref{fig:mlsdss_mass_radius} has a clear separation between active and inactive stars, however most of the active stars have the same locus as some inactive stars. 
To study this overlap in more detail, we divided the sample in different metallicity bins.
We fit the mass-radius relation for active and inactive stars for two different metallicity bins: a low metallicity bin of $0.73\leq\zeta _{\rm TiO/CaH}<0.94$ which corresponds roughly to $-0.5\leq{\rm [Fe/H]} < -0.15$\,dex and a solar metallicity bin of $0.97\leq \zeta _{\rm TiO/CaH}<1.09$ which is approximately $-0.1 \leq {\rm [Fe/H]} < 0.1$\,dex.
Our results are in Figure~\ref{fig:mlsdss_mass_radius_feh}. 
We included the fractional difference between radii of active and inactive stars for both metallicity bins, and we calculated the median of the difference for four mass bins. 
We estimated the uncertainty of this difference by doing a Monte Carlo propagation of uncertainties. 
The median values of fractional difference between radii for active and inactive stars, in order of increasing mass bin are: for low metallicity ${0.005}_{-0.037}^{+0.041}$, ${0.04}_{-0.04}^{+0.03}$, ${0.05}_{-0.03}^{+0.04}$ and ${0.06}_{-0.04}^{+0.04}$, and for solar metallicity ${0.002}_{-0.044}^{+0.031}$, ${0.02}_{-0.03}^{+0.03}$, ${0.05}_{-0.04}^{+0.04}$ and ${0.09}_{-0.08}^{+0.04}$. 
These values of radius inflation are consistent with the measurements from the literature.
By separating in metallicity, the difference between active and inactive stars becomes more clear, and the overlap between the two types of stars is significantly reduced, in comparison to Figure~\ref{fig:mlsdss_mass_radius}. 
Finally, the fractional difference in radius as a function of mass shows that the difference between active and inactive is significant and decreases with mass. 
This difference is not significant for masses $<0.3\,{\rm M_\odot}$ because the uncertainty in our radii calculation becomes comparable to the expected inflation (e.g., $5.8\%$ found by \citealt{Morales2022}).

Besides metallicity, age could also affect the scatter in Figures~\ref{fig:mlsdss_mass_radius} and \ref{fig:mlsdss_mass_radius_feh}. 
M~dwarfs do not change their properties significantly once they converge onto the main sequence. Therefore, a significant difference would only appear for very young stars ($<1$\,Gyr). 
Most of the stars in this sample are within $200$\,pc, which means that if they are young, they belong to a known moving group.
\citet{Kiman2019} showed that these stars needed to be $24$\,Myr old to explain their position in the color-magnitude diagram, which was unlikely. In addition, \citet{Kiman2021} checked the membership of all these stars and found that only 46 stars from this sample belong to moving groups, therefore most of the stars are from the field. From the members of moving groups, most of the stars belong either to the Praesepe cluster or Coma Berenices. By comparing these members of young associations with the active stars, we found that age alone cannot explain the shift in radius between active and inactive stars.

In our work we compared stars that are not young and have similar metallicities, which allows as to identify a difference in radius between active and inactive stars in a large sample. We found that active stars have a larger radius than inactive stars on average, which indicates that the radius inflation problem could be caused by activity. In this work we do not compare to models because the comparison of data to models requires precise ages and metallicities for each star \citep{Torres2018,Torres2019,Torres2020,Torres2021a}, which we do not have for this sample.


\subsection{Radius as a function of $\haew$}

\begin{figure}[ht!]
\begin{center}
\includegraphics[width=\linewidth]{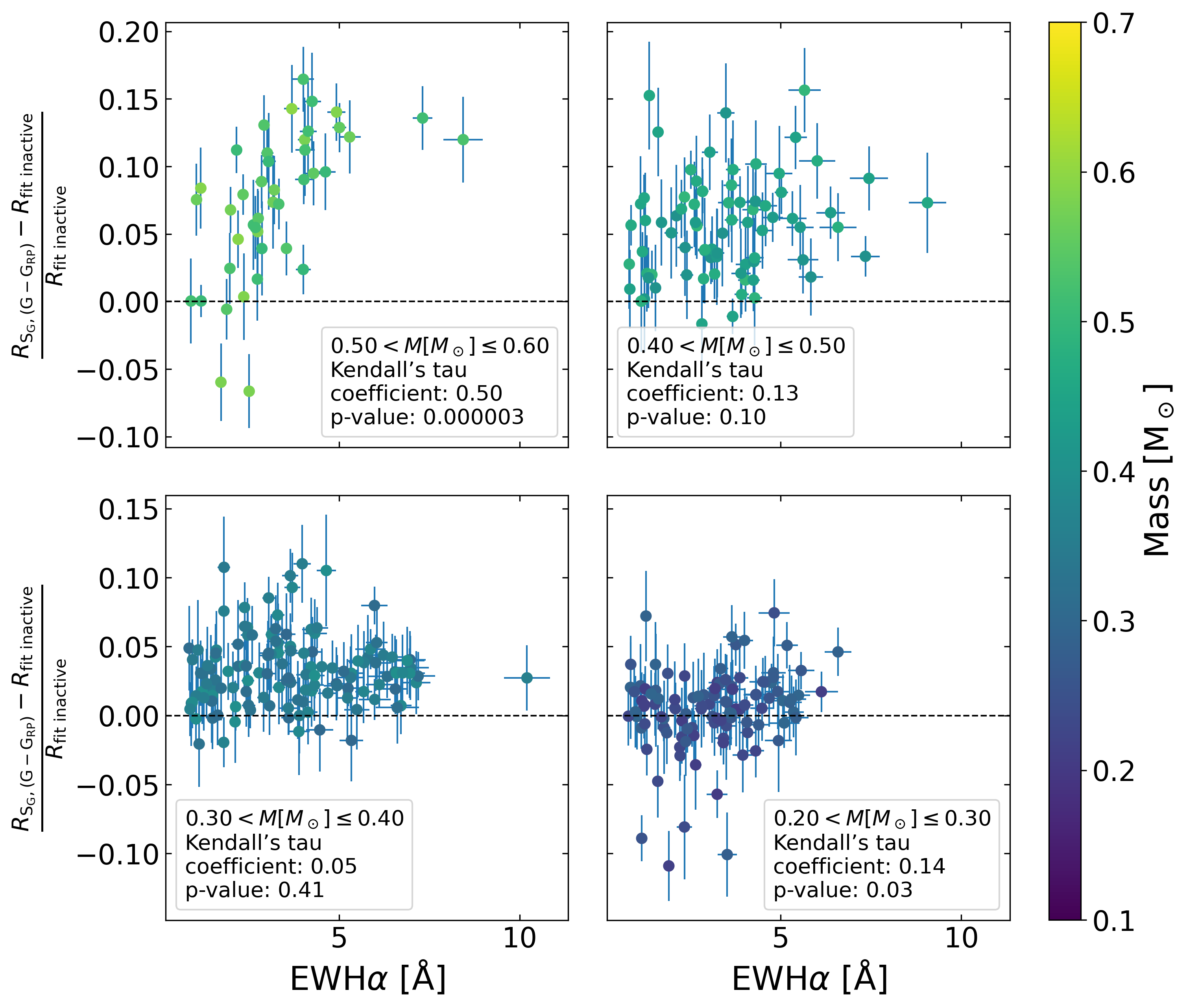}
\caption{Estimated radii for the \citet{Kiman2019} sample of M~dwarfs as a function of $\haew$, divided in four mass bins: $0.5<M[{\rm M_\odot}]\leq 0.6$, $0.4<M[{\rm M_\odot}]\leq 0.5$, $0.3<M[{\rm M_\odot}]\leq 0.4$ and $0.2<M[{\rm M_\odot}]\leq 0.3$. We estimated the Kendall’s $\tau$ coefficient and the p-value to test the hypothesis of no correlation in the data. The results are shown in the label of each panel.} 
\label{fig:mlsdss_radius_vs_halpha}
\end{center}
\end{figure}

The mass-radius relation of active and inactive stars in Figure~\ref{fig:mlsdss_mass_radius} shows that active stars have larger radii than inactive stars. 
In this section we used the same sample from \citet{Kiman2019} to study the dependence of radius inflation on $\haew$ for different mass bins.
We calculated the difference between the radius of active stars and the fit to the mass-radius relation for inactive stars with close to solar metallicity from Figure~\ref{fig:mlsdss_mass_radius_feh}. We also estimated the uncertainties of the difference by doing a Monte Carlo propagation. 
The fractional difference between the active and inactive radii for each active star as a function of $\haew$ is on Figure~\ref{fig:mlsdss_radius_vs_halpha}. 
As the value of $\haew$ for inactive stars does not provide extra information besides that the stars are inactive, we did not include them in the plot.
We divided the stars in four mass bins: $0.5<M[{\rm M_\odot}]\leq 0.6$, $0.4<M[{\rm M_\odot}]\leq 0.5$, $0.3<M[{\rm M_\odot}]\leq 0.4$ and $0.2<M[{\rm M_\odot}]\leq 0.3$.
As in the mass-radius relation in Figure~\ref{fig:mlsdss_mass_radius}, we can see the dependence on mass of the radius inflation, where active higher mass stars have a larger difference with inactive stars than lower mass stars. 
In addition, the top two panels show a correlation between the fractional radius difference and $\haew$. 
We estimated the Kendall’s $\tau$ coefficient and the p-value to test the hypothesis of no correlation in the data\footnote{To calculate the Kendall's $\tau$ coefficient and the p-value we used the python function \texttt{scipy.stats.kendalltau}}. We show the results in the label of each panel of Figure~\ref{fig:mlsdss_mass_radius}. 
As was clear by eye, both bottom panels have low Kendall's $\tau$ coefficients and relatively high p-values, which indicates that there is no evidence for correlation. 
We found similar results for the mass bin $0.4<M[{\rm M_\odot}]\leq 0.5$. 
On the other hand, the panel for the mass bin $0.5<M[{\rm M_\odot}]\leq 0.6$ seems to present a correlation. The p-value for the data in this mass bin is $3\times 10^{-6}$ which means that the two parameters appear substantially more correlated than would be expected by chance.
This indicates that for the mass bin $0.5<M[{\rm M_\odot}]\leq 0.6$, the radius inflation of active stars over inactive ones  is increasing with increasing magnetic activity indicated by $\haew$. 
That there is no clear correlation between radius inflation and $\haew$ on the other three mass bins could be explained by the radius uncertainty of our method ($4\%$). If our uncertainty is similar to or larger than the percentage of radius inflation, then that would not allow us to distinguish a clear trend in the data. 
The lack of correlation could also be physical, however more precise radii measurements are required to distinguish between the two scenarios. 

To ensure that we understand the correlation in Figure~\ref{fig:mlsdss_radius_vs_halpha}, we analyzed each axis separately. The $\haew$ measurements were made from SDSS spectra which has a resolution R$\sim 2000$. It has been shown that $\haew$ increases with decreasing mass for M~dwarfs \citep[e.g., ][]{West2011,Schmidt2015,Kiman2019,Popinchalk2021}. 
Therefore it is important to divide the sample into mass bins, to remove the dependence of $\haew$ with color.
The mass bin size selected in Figure~\ref{fig:mlsdss_radius_vs_halpha} is small enough that there should not be an effect due to the color dependence. 
The masses on this plot were estimated from absolute $K_{\rm s}$ magnitude, meaning that effectively we are selecting a small bin in absolute $K_{\rm s}$ magnitude.
Given that the scatter in the mass-luminosity relation from \citet{Mann2019} is small, we can consider this narrow bin in $K_{\rm s}$ magnitude to be narrow in mass as well.
In addition, the radius estimations using our Gaia SBCR are proportional to the color of the stars (Equation~\eqref{eq:model}); an increase in radius at fixed absolute ${\rm K_{s}}$ implies a redder color. This correlation is expected given that when stars become inflated, their radius increases at constant bolometric luminosity, which generates a decrease in effective temperature and therefore, a redder color. 

The results in Figure~\ref{fig:mlsdss_radius_vs_halpha} partially agree with the results from \citet{Stassun2012}. In that work, the authors studied radius inflation as a function of $\lhalbol$ combining all the masses between $0.3-0.7\,M_\odot$ in the same analysis. 
As we mentioned above, the difference in radius between active and inactive stars is dependent on mass, so it is necessary to study radius inflation in bins of mass to distinguish between dependence on mass and on magnetic activity. 
\citet{Stassun2012} also studied this relation for eclipsing binaries, assuming that the dependence on magnetic activity is the same for eclipsing binaries and single stars. Although we see radius inflation both in single stars and eclipsing binaries, it has been shown that eclipsing binaries tend to rotate faster than single stars of the same mass and age \citep{Torres2021a}. 
Therefore we only compared our analysis to their results on single stars. 

\section{Conclusions}

In this work we developed a method to estimate radii of FGK and M dwarfs using Gaia~DR3 photometry and astrometry. This method has the advantage that it estimates precise radii ($4\%$ average uncertainty) using only photometry and parallaxes provided by the Gaia catalog. Therefore, it is straightforward to estimate radii for large samples.
We used this method to study radius inflation for a statistically large sample of M~dwarfs and showed that radius inflation on single stars is correlated with magnetic activity.

We calibrated the surface brightness-color relation (SBCR) using Gaia~DR3 magnitudes. 
We used the sample of angular diameter measurements from \citet{Duvert2016} as a calibration sample. 
We cross-matched it with Gaia~DR3 to obtain photometry and parallax measurements for the sample. In addition, we searched the literature for metallicity measurements. 
We fit the relation between the flux $S_{\rm m_\lambda}$ defined in Equation~\eqref{eq:SBCR_flux}, color ($X$), and metallicity ([Fe/H]) for each of the three magnitudes in Gaia ($m_\lambda=\{G,G_{\rm RP},G_{\rm BP}\}$) and the three colors ($X=\{(G_{\rm BP}-G_{\rm RP}),(G_{\rm BP}-G),(G-G_{\rm RP})\}$). We used a mixture model with the likelihood shown in Equation~\eqref{eq:likelihood} to fit the relation, and the results are shown in Table~\ref{table:results_fit}. 
We found that our method has an average precision of $4\%$. 

We found good agreement between our Gaia SBCR and two recent calibrations of this relation for the $(G-K_{\rm s})$ color and $S_{\rm G}$ from \citet{Graczyk2021} and \citet{Salsi2021}. However, \citet{Graczyk2021} did not calibrate the relation for masses smaller than $0.6\,\Msun$ and the calibration in \citet{Salsi2021} presents a systematic trend when estimating radii for M~dwarfs which is added by the method. In addition, both SBCRs have a much larger radius uncertainty in comparison to our results for several stars when the uncertainty of the $K_{\rm s}$ magnitude is larger than the uncertainty in Gaia magnitudes. All of these results show the advantage of using our Gaia SBCR. 
We also confirmed the accuracy of our method when we found good agreement between our radius measurements and two sample from the literature: a set of M~dwarfs from \citet{Mann2015} and a sample of FGK and M dwarfs from the Planetary Systems Composite Parameters Table from the NASA Exoplanet Archive. 
We also derived radius with our method for the $50$\,pc sample from Gaia~DR3 which contains model derived radius measurements. We found that the Gaia radius measurements for high mass stars ($>0.6\,\Rsun$) are systematically larger by $3-4\%$ in comparison to our results, and for smaller stars their radii are also larger than our calculations but the offset is as large as $25\%$. We found that this offset is generated by the model-dependent method used in the Gaia catalog.

There are several properties of the star that can affect radius measurements. We studied metallicity, variability and extinction. We found that when using our method to estimate radii, assuming solar metallicity is approximately correct for $-0.3<{\rm [Fe/H]}<0.3$\,dex. Outside that range it is necessary to take into account metallicity. We also found that, given that Gaia averages several photometric measurements, photometric variability is not an issue when estimating radii with Gaia photometry. Finally, we found that the radii estimated using the $(G-G_{\rm RP})$ color are the most affected by extinction for F, G and K, while for M dwarfs $(G_{\rm BP}-G)$ is the most problematic color. 

We studied the influence of magnetic activity on radius measurements --known as radius inflation-- using the sample of M~dwarfs from \citet{Kiman2019}. By comparing field stars of similar metallicity we found that active stars according to $\halpha$ have larger radii than inactive stars. The percentage of inflation decreases with decreasing mass, with a maximum of $9\%$ which agrees with literature measurements. Finally, we used the same sample to study the dependence of radius inflation with $\haew$ for different mass bins. We found a clear correlation for the mass bin $0.5<M[{\rm M_\odot}]\leq 0.6$. 
We found no significant correlation for lower mass stars, although we cannot distinguish if this is due to lack of precision in the radius measurements or to a physical reason. More precise radius measurements are needed to distinguish between the two cases. 
In future work we will study radius inflation using rotation period as magnetic activity indicator. The use of rotation periods will allow us to expand our analysis to higher mass stars.

\section{Acknowledgements}

The authors would like to thank Dr. Tabetha Boyaijan, Dr. Aurora Kesseli, Dr. Mercedes L\'opez-Morales and Dr. Lars Bildsten for helpful discussions that contributed to the results of this work. This project was developed in part at the 2022 NYC Gaia DR3 F\^ete Workshop at the Center for Computational Astrophysics of the Flatiron Institute in 2022 June. This research was supported in part by the National Science Foundation under Grant No. NSF PHY-1748958 and by the Simons Foundation (668346, JPG). This research has made use of the NASA Exoplanet Archive, which is operated by the California Institute of Technology, under contract with the National Aeronautics and Space Administration under the Exoplanet Exploration Program. This work has made use of data from the European Space Agency (ESA) mission {\it Gaia} (\url{https://www.cosmos.esa.int/gaia}), processed by the {\it Gaia} Data Processing and Analysis Consortium (DPAC,
\url{https://www.cosmos.esa.int/web/gaia/dpac/consortium}). Funding for the DPAC has been provided by national institutions, in particular the institutions participating in the {\it Gaia} Multilateral Agreement.

\software{\texttt{scipy} \citep{SciPy-NMeth2020}; \texttt{numpy} \citep{harris2020array}; \texttt{matplotlib} \citep{Hunter2007}; \texttt{astropy} \citep{astropy2013,astropy2018,astropy2022}; \texttt{emcee} \citep{Foreman-Mackey2013}}

\facility{Exoplanet Archive.}







\bibliographystyle{aasjournal}
\bibliography{bibliography,citation_nasa_archive_rad,citation_nasa_archive_met}

\end{document}